\documentclass[10pt,conference]{IEEEtran}
\usepackage{cite}
\usepackage{amsmath,amssymb,amsfonts}
\usepackage{graphicx}
\usepackage{textcomp}
\usepackage{xcolor}
\usepackage[hyphens]{url}
\usepackage[ruled,vlined,linesnumbered]{algorithm2e}
\usepackage{booktabs}  
\usepackage{multirow,makecell}
\usepackage{pifont} 
\usepackage{enumitem}
\usepackage{subfigure}
\usepackage{wrapfig}

\usepackage{CJK}
\usepackage{type1cm}
\usepackage{times}
\usepackage[marginal]{footmisc}

\usepackage{etoolbox}
\makeatletter
\patchcmd{\@makecaption}
  {\scshape}
  {}
  {}
  {}
\makeatletter
\patchcmd{\@makecaption}
  {\\}
  {.\ }
  {}
  {}
\makeatother

\usepackage{enumitem}

\setenumerate[1]{itemsep=0.5pt,partopsep=0pt,parsep=\parskip, topsep=2pt, leftmargin=12pt}

\newcommand{\M}{HeatViT}

\usepackage{colortbl}  

\def\BibTeX{{\rm B\kern-.05em{\sc i\kern-.025em b}\kern-.08em
    T\kern-.1667em\lower.7ex\hbox{E}\kern-.125emX}}

\title{\M: \underline{H}ardware-\underline{E}fficient \underline{A}daptive \underline{T}oken Pruning for \underline{Vi}sion \underline{T}ransformers} 
\author{
    Peiyan Dong\textsuperscript{\rm 1},\ \ \ 
    Mengshu Sun\textsuperscript{\rm 1},\ \ \ \
    Alec Lu\textsuperscript{\rm 2}, \ \
    Yanyue Xie\textsuperscript{\rm 1}, \ \
    Kenneth Liu\textsuperscript{\rm 2}, \ \
    Zhenglun Kong\textsuperscript{\rm 1}, \\
    Xin Meng\textsuperscript{\rm 1}, \ \
    Zhengang Li\textsuperscript{\rm 1}, \ \
    Xue Lin\textsuperscript{\rm 1}, \ \
    Zhenman Fang\textsuperscript{\rm 2}, \ \
    Yanzhi Wang\textsuperscript{\rm 1,3}\\
    \small
    \textsuperscript{1}Northeastern University,\quad
    \textsuperscript{2}Simon Fraser University,\quad
    \textsuperscript{3}CoCoPIE LLC\\
    {\tt\small \{dong.pe, sun.meng, xie.yany, kong.zhe, li.zhen, xue.lin, yanz.wang\}@northeastern.edu,}\\
    {\tt\small \{alec\_lu, ksl24, zhenman\}@sfu.ca,}
    {\tt\small 1601214372@pku.edu.cn}
}

\begin{document}
\maketitle
\thispagestyle{plain}
\pagestyle{plain}


\begin{abstract}
While vision transformers (ViTs) have continuously achieved new milestones in the field of computer vision, their sophisticated network architectures with high computation and memory costs have impeded their deployment on resource-limited edge devices.
In this paper, we propose a hardware-efficient image-adaptive token pruning framework called \textbf{\M} for efficient yet accurate ViT acceleration on embedded FPGAs.
By analyzing the inherent computational patterns in ViTs, we first design an effective attention-based multi-head token selector, which can be progressively inserted before transformer blocks to dynamically identify and consolidate the non-informative tokens from input images.
Moreover, we implement the token selector on hardware by adding miniature control logic to heavily reuse existing hardware components built for the backbone ViT.
To improve the hardware efficiency, we further employ 8-bit fixed-point quantization and propose polynomial approximations with regularization effect on quantization error for the frequently used nonlinear functions in ViTs.
Finally, we propose a latency-aware multi-stage training strategy to determine the transformer blocks for inserting token selectors and optimize the desired (average) pruning rates for inserted token selectors, in order to improve both the model accuracy and inference latency on hardware.
Compared to existing ViT pruning studies, under the similar computation cost, \M~can achieve 0.7\%$\sim$8.9\% higher accuracy; while under the similar model accuracy, \M~can achieve more than 28.4\%$\sim$65.3\% computation reduction, for various widely used ViTs, including DeiT-T, DeiT-S, DeiT-B, LV-ViT-S, and LV-ViT-M, on the ImageNet dataset.
Compared to the baseline hardware accelerator, our implementations of \M~on the Xilinx ZCU102 FPGA achieve $3.46\times$$\sim$$4.89\times$ speedup with a trivial resource utilization overhead of 8\%$\sim$11\% more DSPs and 5\%$\sim$8\% more LUTs.\\ \par\
{\bf\emph{Keywords- Vision Transformer; FPGA Accelerator; Hardware and Software Co-design; Data-level Sparsity.}\rm}
\end{abstract}

\section{Introduction}
\label{sec:introduction}

Transformers \cite{bahdanau2015neural,parikh2016decomposable} have recently made an attractive resurgence in the form of Vision Transformers ({ViTs}) \cite{dosovitskiy2020image}, showing strong versatility in NLP \cite{vaswani2017attention,kenton2019bert,brown2020language}, computer vision (e.g., image classification \cite{dosovitskiy2020image}, object detection \cite{carion2020end,zhu2020deformable}, semantic segmentation \cite{zheng2021rethinking}, image processing \cite{chen2021pre}, and video understanding \cite{zhou2018end}), and complex scenarios with multi-modal data.  
Furthermore, ViTs can be used as effective backbone networks \cite{dosovitskiy2020image,wang2021pyramid,han2021transformer,cao2021swin} with superior transferability to downstream tasks through minor fine-tunings.
Seemingly, ViTs and transformers have great potential to unify diverse application domains through common architectures and tackle the reliance on scarce domain data,  ultimately addressing the two fundamental problems in deep learning: (i) strong reliance on domain data, and (ii) constant model improvements to serve evolving needs.
ViTs and transformers are largely considered as one of the future dominant deep learning techniques.

However, to fully unleash advantages of transformer architectures, we need to address the following \emph{challenges} before ViTs and transformers become an indispensable staple in future AI computing.
(i) Although the self-attention mechanism is a key defining feature of transformer architectures, a well-known concern is its quadratic time and memory complexity with respect to the number of input tokens. This hinders scalability in many settings, let alone deployment on resource-constrained edge devices.
(ii) Majority of existing works on efficient ViT and transformer techniques followed what has been done for Convolutional Neural Networks (CNNs) by using conventional weight pruning \cite{zhu2021visual, chen2021chasing, yu2022unified, yu2021unified}, quantization \cite{zhao2020investigation,prato2020fully,liu2021post}, and compact architecture design \cite{guo2019nat,so2019evolved,wang2020hat,chen2021glit,chen2021autoformer,gong2021nasvit,chen2021searching,li2021bossnas,wu2021cvt}, with limited accuracy and speed performance.
(iii) In the efforts of exploring token removal to address the quadratic complexity, the static approaches \cite{rao2021dynamicvit, liang2022evit,fayyaz2021, tang2021patch, chen2021chasing} remove tokens with a fixed ratio in an input-agnostic manner, ignoring input sample dependent redundancy; and existing image-adaptive approaches \cite{pan2021iared2, yu2022unified, xu2021evo} simply discard non-informative tokens and do not fully explore token-level redundancies from different attention heads.
Both approaches achieve a relatively low pruning rate (to preserve accuracy) or an undermined accuracy (at a high pruning rate).
Moreover, none of these studies support efficient hardware implementation on edge devices.
(iv) Transformer architectures tend to use more hardware-unfriendly computations, e.g., more nonlinear operations than CNNs, to improve accuracy.
Therefore, we need to address the low hardware efficiency issue of such computations, while enjoying the additional optimization dimension provided by multi-head self-attention.

\begin{figure}[tb]
\centering
\includegraphics[width=0.95\columnwidth]{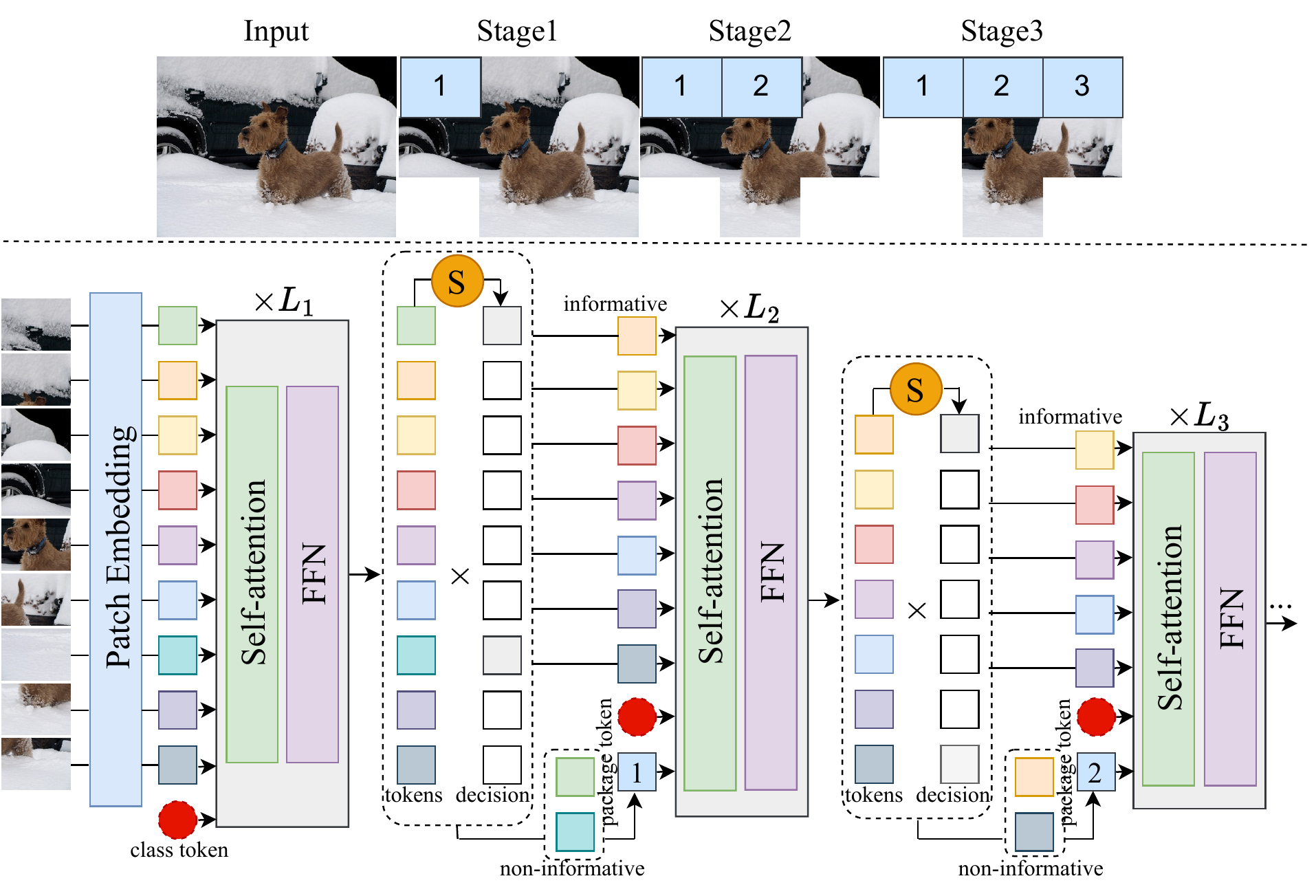}
\vspace{-0.1in}\caption{The workflow of \M~on DeiT-S. The proposed token selector S conserves informative tokens conditioned on features from the previous layer. And non-informative tokens are averaged into one package token (blue).}
\label{fig:workflow}
\vspace{-0.2in}
\end{figure}

In this paper, we propose \M, a hardware-efficient image-adaptive token pruning framework, together with 8-bit quantization, for efficient yet accurate ViT inference acceleration on embedded FPGA platforms. To improve pruning rate while preserving model accuracy, we make two observations by analyzing the computation workflow in ViTs: (i) the information redundancy in input tokens differs among attention heads in ViTs; and (ii) non-informative tokens identified in earlier transformer blocks may still encode important information when propagating to later blocks.
Based on these, we design an effective token selector module that can be inserted before transformer blocks to reduce token number (i.e., number of tokens) with negligible computational overhead.
As shown in Fig.~\ref{fig:workflow}, we incorporate different token-level redundancies from multiple attention heads to be more accurate in token scoring.
Furthermore, instead of completely discarding non-informative ones, we package them together into one informative token to preserve information for later transformer blocks.

For hardware implementation on edge FPGA devices, we design our token selector with linear layers, i.e., fully-connected (FC) layers, instead of convolutional (CONV) layers, to reuse the GEMM (General Matrix Multiply) hardware component built for the backbone ViT (i.e., without token selectors) execution. 
In addition, we always concatenate the identified (and sparse) informative tokens and the packaged informative token together to form a dense input of tokens to avoid sparse computations on hardware. 
To improve the hardware efficiency, we further apply 8-bit quantization for weights and activations, and propose polynomial approximations for the frequently used nonlinear functions in ViTs, including GELU~\cite{hendrycks2016gaussian}, Softmax~\cite{Goodfellow-et-al-2016}, and Sigmoid~\cite{mitchell1999machine}. 
Besides, we introduce the regularization effect on quantization error into the design of polynomial approximations to support more ambitious quantization.
We design a proof-of-concept FPGA accelerator for ViTs based on our proposed HeatViT. 
We implement the GEMM engine inspired by~\cite{FPT19gemm,lizh2022logic} to execute the most computation-intensive multi-head self-attention module and feed-forward network in the backbone ViT, and the classification network in our token selector. 
It is worth noting that only lightweight control logic is added to support our adaptive token pruning by reusing the same hardware components built for the backbone ViT execution.

To reduce inference latency on hardware while preserving model accuracy (typically within 0.5\% or 1\% accuracy loss), we propose a latency-aware multi-stage 
training strategy to (i) determine the transformer blocks for inserting token selectors and (ii) optimize the desired (average) pruning rates for these token selectors.
Specifically, we insert a token selector for each transformer block, from the last block backward to the first one, since early blocks are more sensitive to accuracy. 
For each insertion, we train the current token selector while fine-tuning other network components, to decrease latency through increasing pruning rate of the current block until accuracy loss exceeds a threshold. 
Then we further consolidate token selectors in consecutive transformer blocks with similar pruning rates into one token selector for a whole stage of transformer blocks.
Our training strategy is conducted as fine-tuning on pre-trained ViTs.
And because we use small epoch numbers for inserting token selectors, our training effort is roughly 90\% of that for training-from-scratch of backbone ViTs (without token selectors).


\begin{table}[htb]
\centering
\tabcolsep 5pt
\vspace{-0.1in}
\caption{Comparison between representative ViT pruning methods.}
\vspace{-0.1in}\label{tab:pruning_works}
\begin{tabular}{c|c|c|c}
\toprule
\textbf{Method}                      & \textbf{\makecell{Design \\ Scheme}} & \textbf{Method} & \textbf{\makecell{Design \\ Scheme}}  \\
\midrule
DyViT~\cite{rao2021dynamicvit}        & \ding{172}                          & ATS~\cite{fayyaz2021}      & \ding{172}           \\ \hline
IA        ~\cite{pan2021iared2} 	  & \ding{173}                          & PS-ViT~\cite{tang2021patch}& \ding{172}           \\ \hline
VTP~\cite{zhu2021visual}              & \ding{174}                          & Evo-ViT~\cite{xu2021evo}   & \ding{173}          \\ \hline
S2ViTE~\cite{chen2021chasing}         & \ding{172}\ding{174}\ding{175} 		& UP-DeiT~\cite{yu2021unified}& \ding{174}\ding{175} \\ \bottomrule
\end{tabular}
{\\ \footnotesize \raggedright \ding{172} -- Static Token Pruning; \ding{173} -- Adaptive Token Pruning; \ding{174} -- Head Pruning; \\ \ding{175} -- Token Channel Pruning.\par}
\vspace{-0.2in}
\end{table}

\begin{figure*}[htb]
\centering
\includegraphics[width=1.2\columnwidth]{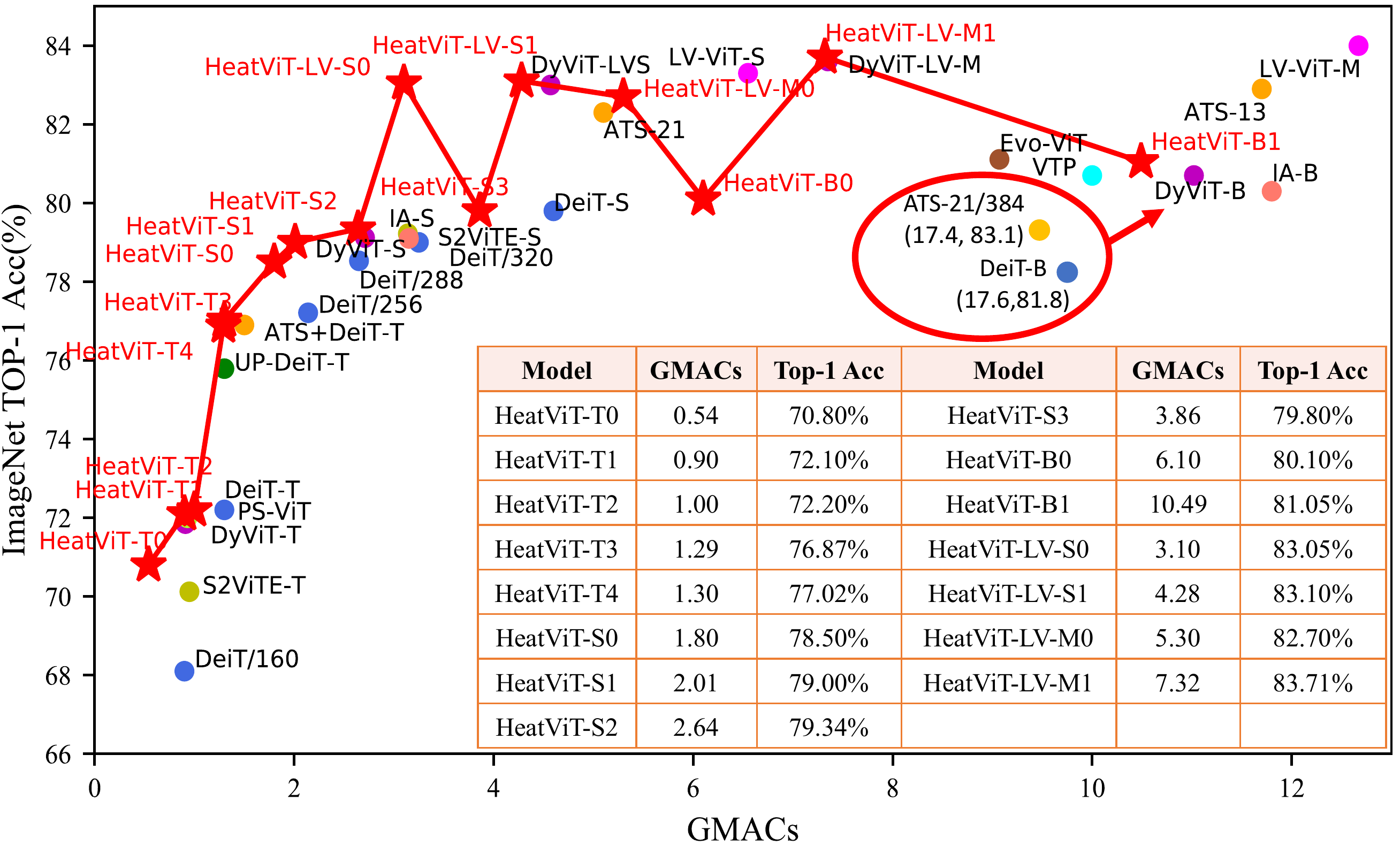}
\vspace{-8pt}\caption{Comparison of \M~with prior pruning methods: Under a similar computation cost, \M~achieves 0.7\%$\sim$8.9\% higher accuracy; while under a similar accuracy, \M~reduces the computation cost by more than 28.4\%$\sim$ 65.3\%. Final \M~models are quantized into 8-bit fixed-point format.}
\label{fig:user-trade-off}
\vspace{-0.4cm}
\end{figure*}

As summarized in Fig.~\ref{fig:user-trade-off}, we evaluate \M~for multiple widely used ViT models on the ImageNet dataset, including DeiT-tiny (DeiT-T, HeatViT-T), DeiT-small (DeiT-S, HeatViT-S), DeiT-base (DeiT-B, HeatViT-B)~\cite{Touvron2021TrainingDI}, LV-ViT-small (LV-ViT-S, HeatViT-LV-S), and LV-ViT-medium (LV-ViT-M, HeatViT-LV-M)~\cite{jiang2021all}. 
These models are already well condensed and thus more challenging to prune compared to larger ViT models. 
Compared to state-of-the-art ViT pruning studies in Table~\ref{tab:pruning_works} (pruning types are explained in detail in Sec~\ref{sec:pruning_vit}), HeatViT achieves better accuracy-computation trade-offs: 
Under a similar computation cost, \M~can achieve 0.7\%$\sim$8.9\% higher accuracy; while under a similar accuracy, \M~can reduce the computation cost by more than 28.4\%$\sim$ 65.3\%.
Compared to the baseline hardware accelerator (16-bit and no token pruning), our implementations of \M~on the Xilinx ZCU102 FPGA achieves $3.46\times$$\sim$$4.89\times$speedup with a trivial resource utilization overhead of 8\%$\sim$11\% more DSPs and 5\%$\sim$8\% more LUTs.
Compared to the optimized ARM CPU and NVIDIA GPU versions on the NVIDIA Jetson TX2 board with our token pruning, our FPGA implementations achieve 685$\times$$\sim$1695$\times$ and 2.68$\times$$\sim$3.79$\times$ more speedups, 242.6$\times$$\sim$719.0$\times$ and 3.0$\times$$\sim$4.7$\times$ higher energy efficiency.


Our contributions are summarized as follows:
\begin{itemize}[leftmargin=*]

    \item Algorithm and hardware co-design for an effective and hardware-efficient token selector to enable efficient image-adaptive token pruning in ViTs.
    \item A latency-aware multi-stage training strategy to learn the effective insertion of token selectors in ViTs.
    \item A polynomial approximation of nonlinear functions inside ViTs for more ambitious quantization and efficient FPGA-based implementations.
    \item An end-to-end acceleration framework, with both image-adaptive pruning and 8-bit quantization, for ViT inference on embedded FPGAs.
    \item Experiments to demonstrate superior pruning rates and inference accuracy of \M~over state-of-the-art ViT pruning studies, as well as trivial hardware resource overhead.

\end{itemize}

\begin{figure}[tb]
\centering
\includegraphics[width=0.8\columnwidth]{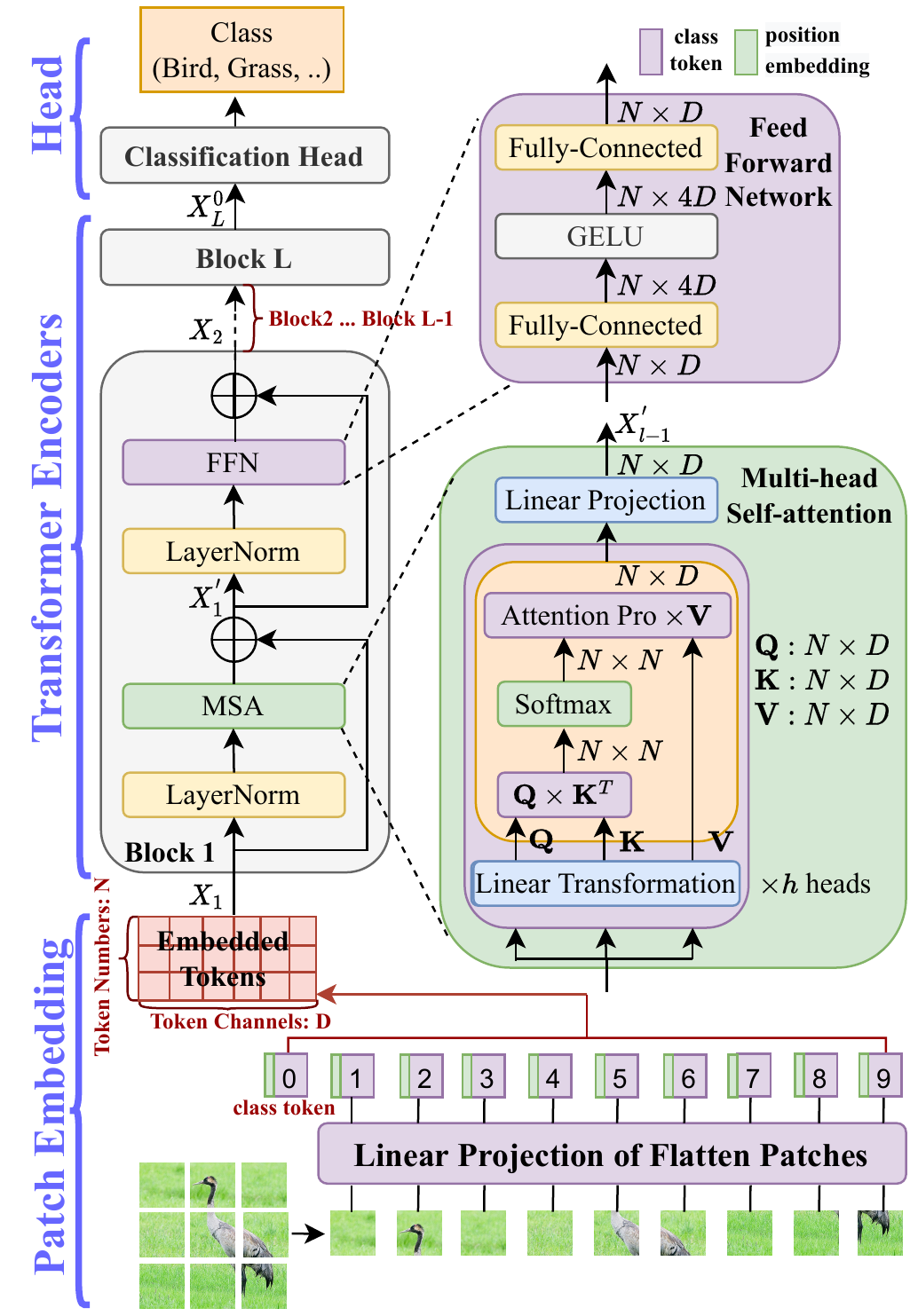}
\vspace{-0.1in}
\caption{Overview of ViT}
\label{fig:encoder}
\vspace{-0.3in}
\end{figure}
\section{Background and Related Work}
\label{sec:related_work}

\subsection{Vision Transformer}
Transformers are initially proposed to handle the learning of long sequences in NLP tasks. Dosovitskiy et al.~\cite{dosovitskiy2020image} and Carion et al.~\cite{carion2020end} adapt the transformer architecture to image classification and detection, respectively, and achieve competitive performance against CNN counterparts with stronger training techniques and larger-scale datasets. DeiT~\cite{Touvron2021TrainingDI} further improves the training pipeline with the aid of distillation, eliminating the need for large-scale pretraining~\cite{yuan2021tokens}. Inspired by the competitive performance and global receptive field of transformer models, follow-up works design ViTs for different computer vision tasks including object detection, semantic segmentation, 3D object animation, and image retrieval.

Fig.~\ref{fig:encoder} illustrates three components in ViTs: \emph{patch embedding}, \emph{transformer encoders}, and \emph{classification output head}.
For \textbf{patch embedding}, an image $\mathbf X \in \mathbb R^{H \times W \times C}$ is  reshaped into a sequence of 2D patches $\mathbf X_p \in \mathbb R^{N \times (P^2 \cdot C)}$, where $(H,W)$ is  original image resolution, $C$ is channel number, $(P,P)$ is the resolution of  patches, and $N=HW/P^2$ is the number of patches and also the effective input sequence length (i.e., number of tokens) to transformer encoders.
Patch embeddings are obtained by mapping $\mathbf X_p$ into $D$ dimensions via a trainable linear projection $\mathbf E \in \mathbb R^{(P^2 \cdot C) \times D}$: $\mathbf X_0 =$ $[\mathbf X_{class}; \mathbf X^1_p \mathbf E; \mathbf X^2_p \mathbf E; ...; \mathbf X^N_p \mathbf E]+\mathbf E_{pos}$,
where $D$ is the constant latent vector size throughout transformer encoders, and $\mathbf E_{pos} \in \mathbb R^{(N+1) \times D}$ is {learnable} position embeddings. 
A learnable embedding $\mathbf X_0^0 = \mathbf X_{class}$ is prepended in patch embeddings $\mathbf X_0$.
Then $\mathbf X$ goes through a stack of $L$ transformer encoders for processing.
The output $\mathbf X_L^0$ (class token) of the final {($L$-th) transformer encoder}  is fed to \textbf{a classification output head} implemented by an MLP to obtain the \textit{final result}. The MLP module uses two linear layers with a Gaussian Error Linear Unit (GELU) activation layer in between.

The $l$-th \textbf{transformer encoder} receives the patch embeddings $\mathbf X_{l-1}$ as input.
Fig.~\ref{fig:encoder} illustrates a ViT encoder block, consisting of a multi-head self-attention (MSA) module and an MLP (FFN) module, both with LN (layer normalization) applied before input.
The encoder block operations are then:
\begin{equation}
\begin{aligned}
     \mathbf{X'}_{l-1}   &=  \text{MSA} ( \text{LN} ( \mathbf{X}_{l-1} ) ) +  \mathbf{X}_{l-1}, \\
     \mathbf{X}_{l}  &=  \text{FFN} ( \text{LN} (\mathbf{X'}_{l-1})) + \mathbf{X'}_{l-1}.
\end{aligned}
\end{equation}
For $h$ heads in MSA, $\text{LN}(\mathbf{X}_{l-1})$ is split into $h$ parts.
For each head, the corresponding part in $\text{LN}(\mathbf{X}_{l-1})$ is transferred into query $\mathbf{Q}$, key $\mathbf{K}$, and value $\mathbf{V}$ through three linear projections $W_Q$, $W_K$, and $W_V$, respectively. 
Then self-attention function~\cite{vaswani2017attention} in each head is performed as: 
\begin{equation}
     \text{Attention}(\mathbf{Q},\mathbf{K}, \mathbf{V})=\text{Softmax}(\mathbf{Q}\mathbf{K}^T/\sqrt{D})\mathbf{V}.\label{eq:selfattention}
\end{equation}
Attention outputs from all the heads are concatenated and linearly transformed.

\vspace{-5pt}
\subsection{Model Pruning on ViT}\label{sec:pruning_vit}
State-of-the-art ViT pruning studies exploring the redundancy from three dimensions in ViTs are summarized below.

\noindent\textbf{Redundancy of the attention head.}
Attention head pruning~\cite{zhu2021visual, chen2021chasing, yu2022unified, yu2021unified} reduces weight redundancy on the transformation matrices ($W_Q$, $W_K$, $W_V$) before MSA operation. 
Due to the parallel computing nature of the transformer heads, these works directly prune some heads entirely and lead to significant accuracy drop,
for example, 27\% GMACs reduction, but with 2.2\% accuracy drop on DeiT-T in~\cite{chen2021chasing}.
It is an inefficient way of computation reduction because the attention head usually contributes less than 43\% of the total computation in several ViT architectures~\cite{kong2021spvit}.

\noindent\textbf{Redundancy of the token channel.}
Token channel~\cite{chen2021chasing, yu2022unified, yu2021unified} pruning executes feature-level pruning on the embedding of the token feature to diminish the redundancy of the feature representation. 
Since this pruning type is equivalent to structured pruning on the token embedding, i.e., removing the same embedding dimension for different tokens, it is difficult to guarantee an ideal pruning rate without significant accuracy deterioration.

\noindent\textbf{Redundancy of the token number.}
Token pruning~\cite{rao2021dynamicvit, liang2022evit, pan2021iared2, fayyaz2021, tang2021patch, chen2021chasing, xu2021evo, yu2022unified} aims at removing the non-informative input data.
According to~\cite{yang2021instance}, the accuracy of neural networks is more related to the object information, but not the background information, implying the input-level redundancy. There is often much higher redundancy along the token number dimension.

\subsection{Computational Complexity Analysis}\label{sec:computation_complexity}
Given an input sequence $N{\times} D$, where $N$ is the input sequence length or the token number, and $D$ is the embedding dimension~\cite{Touvron2021TrainingDI} of each token,
some works~\cite{pan2021iared2,zhu2021visual} address the computational complexity of ViT as $(12ND^2 + 2N^2D)$.
However, $D$ represents three physical dimensions: (i) $D_{ch}$ -- channel size of a token, (ii)
$h$ -- number of heads, and (iii) $D_{attn_{s}}$ -- sub-channel size for one head. 
The channel size of the Query ($\mathbf{Q}$), Key ($\mathbf{K}$), and Value ($\mathbf{V}$) matrices is $HD_{attn_{s}}$.
As shown in Table~\ref{tab:complexity}, the computational complexity of the ViT can be written as $[4ND_{ch}(hD_{attn_{s}}) + 2N^2(hD_{attn_{s}}) + 8ND_{ch}D_{fc}]$.
Compared with other prunable dimensions, directly pruning tokens, $N$, will contribute to the reduction of all operations linearly or even quadratically ($N^2$ in layers \ding{173} and \ding{174}), more compression benefits than other pruning types.
Please note that since token pruning is equivalent to removing information redundancy within the image data, it can be more freely integrated into the compression process on other model dimensions (e.g., model quantization in our case).

\begin{table}[htb]
\centering
\tabcolsep 3pt

\vspace{-10pt}
\caption{Computational complexity of one ViT block.}
\label{tab:complexity}

\vspace{-5pt}
\scalebox{0.9}{
\begin{tabular}{c | c | c | c | c | c }
\toprule
Layer & Module & Input Size & Computation & Output Size & \#MACs \\ \midrule
\ding{172} & \multirow{5}{*}{MSA} & $N \times D_{ch}$ & \makecell{Linear \\ Transformation} & $N \times hD_{attn_{s}}$ & $ND_{ch}hD_{attn_{s}}$ \\ \cmidrule{1-1} \cmidrule{3-6}
\ding{173} & & $N \times hD_{attn_{s}}$ & $\mathbf{Q} \times \mathbf{K}^\mathrm{T}$ & $N \times N$ & $N^{2}hD_{attn_{s}}$ \\ \cmidrule{1-1} \cmidrule{3-6}
\ding{174} & & $N \times N$ & $\mathbf{QK}^\mathrm{T} \times \mathbf{V}$ & $N \times hD_{attn_{s}}$ & $N^{2}hD_{attn_{s}}$ \\ \cmidrule{1-1} \cmidrule{3-6}
\ding{175} &  & $N \times hD_{attn_{s}}$ & Projection & $N \times D_{ch}$ & $NhD_{attn_{s}}D_{ch}$ \\ \midrule
\ding{176} & \multirow{2}{*}{FFN} & $N \times D_{ch}$ & FC Layer & $N \times 4D_{fc}$ & $4ND_{ch}D_{fc}$  \\ \cmidrule{1-1} \cmidrule{3-6}
\ding{177} & & $N \times 4D_{fc}$ & FC Layer & $N \times D_{ch}$ & $4ND_{fc}D_{ch}$ \\ \midrule
\multicolumn{6}{c}{Total MACs: $4ND_{ch}hD_{attn_{s}} + 2N^{2}hD_{attn_{s}} + 8ND_{ch}D_{fc}$} \\
\bottomrule
\end{tabular}
}
\end{table}

\subsection{Static vs. Image-Adaptive Token Pruning}\label{sec:pruning_vs}
\begin{figure}[tb]
\centering
\includegraphics[width=1\columnwidth]{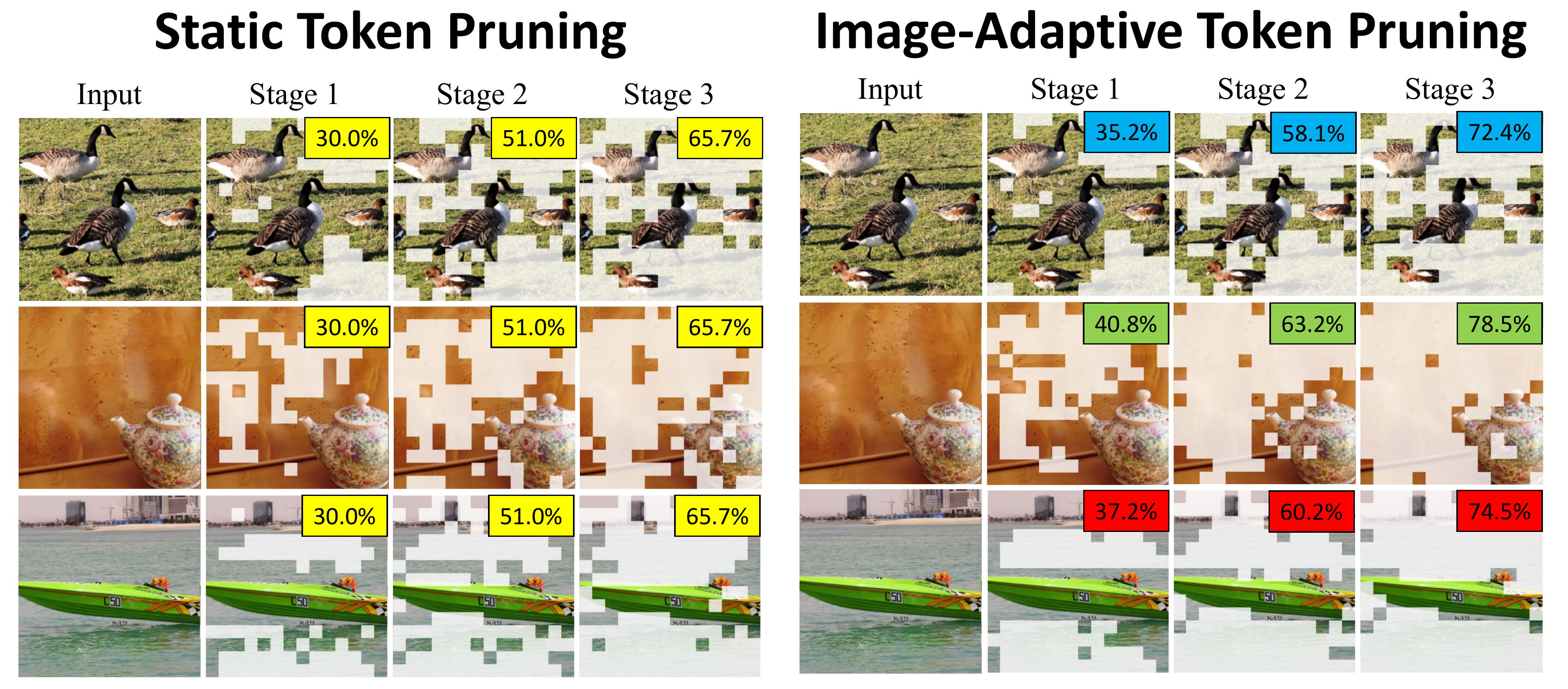}
\vspace{-0.25in}
\caption{Comparison between static token pruning and \M~image-adaptive token pruning.}
\label{fig:token_pruning}
\vspace{-0.2in}
\end{figure}

There are two sub-branches in the token pruning of the ViT: static token pruning and image-adaptive token pruning.
Static token pruning~\cite{rao2021dynamicvit, liang2022evit,fayyaz2021, chen2021chasing, tang2021patch} reduces the number of input tokens by a fixed ratio for different images, which neglects the fact that the information of each image varies both in the region size and location, and thus restricts the image pruning rate.
In contrast, image-adaptive token pruning~\cite{pan2021iared2, yu2022unified, xu2021evo} deletes redundant tokens based on the inherent image characteristics to achieve a per-image adaptive pruning rate.
Generally, smaller pruning rates are applied for complex and information-rich images while larger pruning rates for simple and information-less ones.
Ideally, it can achieve a larger overall pruning rate than the static one as Fig~\ref{fig:token_pruning}.
However, 
existing adaptive token pruning studies do not consider the visual characteristics inside the ViT (token redundancy in ViT, Sec~\ref{sec:motivation}) --visual receptive field of different heads, and delete the less informative tokens completely without the opportunity to correct evaluation errors.
Moreover, there are no more details on how to determine the location, number, and pruning rate of the pruning evaluation module specifically.
Hence, they often lead to a limited overall pruning rate (e.g., 35\% GMACs reduction on DeiT-S~\cite{pan2021iared2}).
Finally, none of these studies have considered the efficient hardware implementation on edge devices to support adaptive token pruning. For example,~\cite{pan2021iared2} introduces a new operation, Argsort, which is currently not compatible with many frameworks~\cite{prillo2020softsort}.

\subsection{Transformer Accelerators on FPGAs}\label{sec:transformerFPGA}
Previous state-of-the-art FPGA accelerators on Transformer-based models \cite{li2020ftrans}, \cite{qi2021accommodating} and \cite{zhang2021algorithm} also leverage sparsity.
However, they have three main limitations.
(i)\cite{li2020ftrans} depends on block-circulant matrix-based weight representation in order to replace the matrix-vector multiplication in FC layers with FFT/IFFT-based processing elements. However,~\cite{li2020ftrans} accelerates only the MSA part of the model and omits the FFN part (about $65\%$ computation of the whole model).
(ii) \cite{qi2021accommodating} utilizes the block-wise weight pruning and \cite{zhang2021algorithm} implements the structure pruning (row-wise or column-wise) to compress weights to similar sizes (lower memory footprint). However, the pruning granularity of both is coarse, resulting in severe accuracy degradation at limited compression rates. For example,~\cite{qi2021accommodating} shows 2\% accuracy drops under the 36\% sparsity.
(iii) These FPGA-based accelerators are mainly for transformers in the Natural Language Processing (NLP) tasks, not for computer vision ones.
In addition, they all compress the model weights, which means that new substructures or additional overhead is needed to support the specific sparse matrix multiplication.
According to our Section~\ref{sec:motivation} analysis, the difference in input data type will introduce a unique redundancy that motivates an effective compression space.
Experiments show that adaptive token pruning at the data level can speed up the model inference with negligible accuracy losses and little overhead for hardware implementations.

\section{Analysis of Token Redundancy in ViT and Overview of \M}
\label{sec:motivation}

\subsection{Token Redundancy in ViT} \label{sec:analysis_token_packaging}
\begin{figure}[tb]
\centering
\vspace{-0.2in}
\includegraphics[width=0.8\columnwidth]{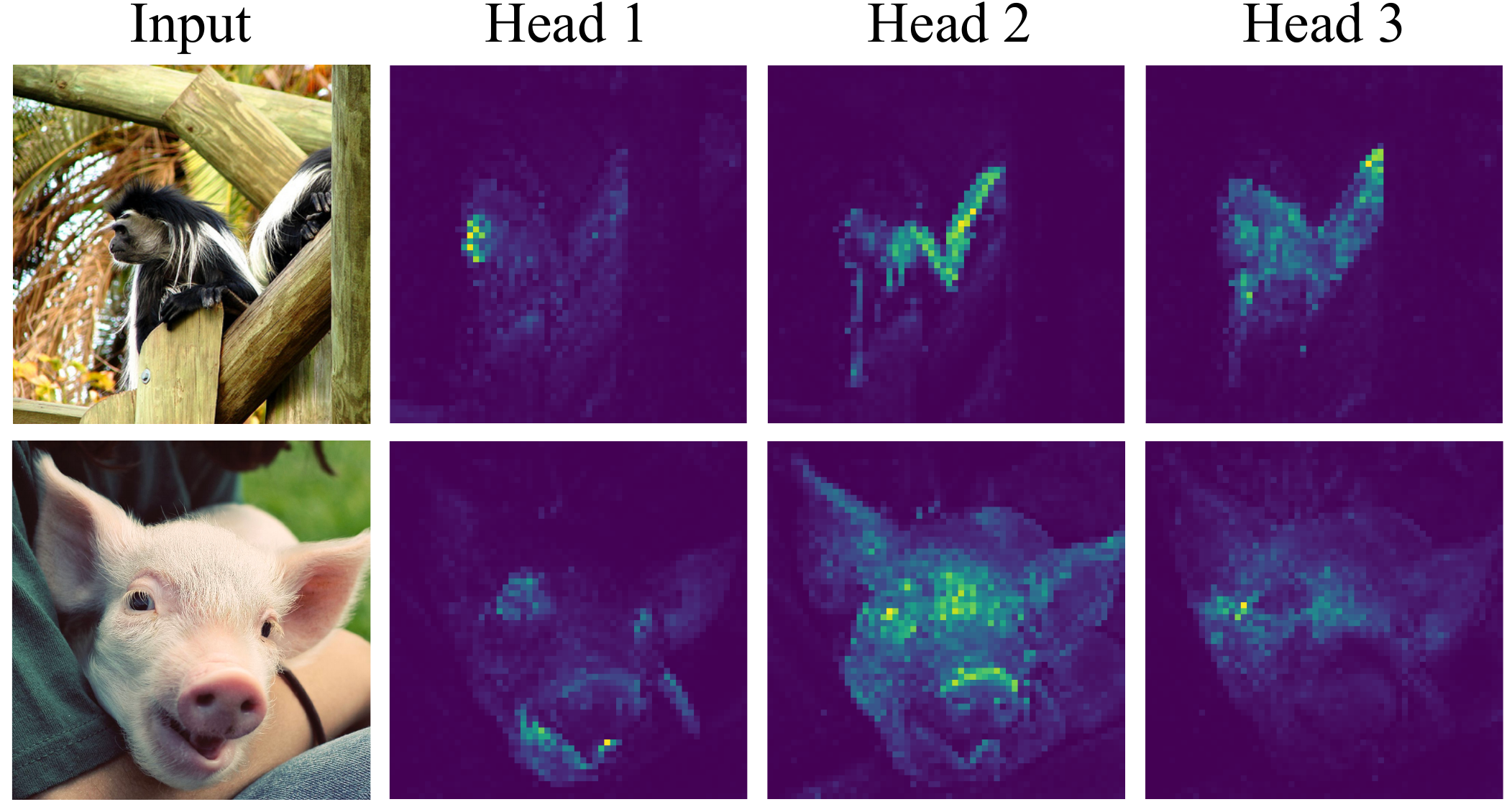}
\caption{The information region detected by each head in DeiT-T.}
\label{fig:head}
\vspace{-0.25in}
\end{figure}

\noindent\textbf{Token redundancy viewed by different attention heads.}
Assuming that the final class (CLS) token is strongly correlated with classification~\cite{zhou2021refiner}, we visualize the information regions detected by the CLS token in different attention heads of DeiT-T, as shown in Fig.~\ref{fig:head}.
It can be seen that each head extracts the features of the image (multi-head visual receptive area of the image) independently and differently~\cite{pan2021iared2,heo2021pit,mao2021dual}, indicating that the heads contain distinct token-level redundancy. This inspires our token selector design with multiple heads, which is elaborated in Section~\ref{sec:token_selector}.

\begin{figure}[tb]
\centering
\vspace{-0.1in}
\includegraphics[width=0.8 \columnwidth]{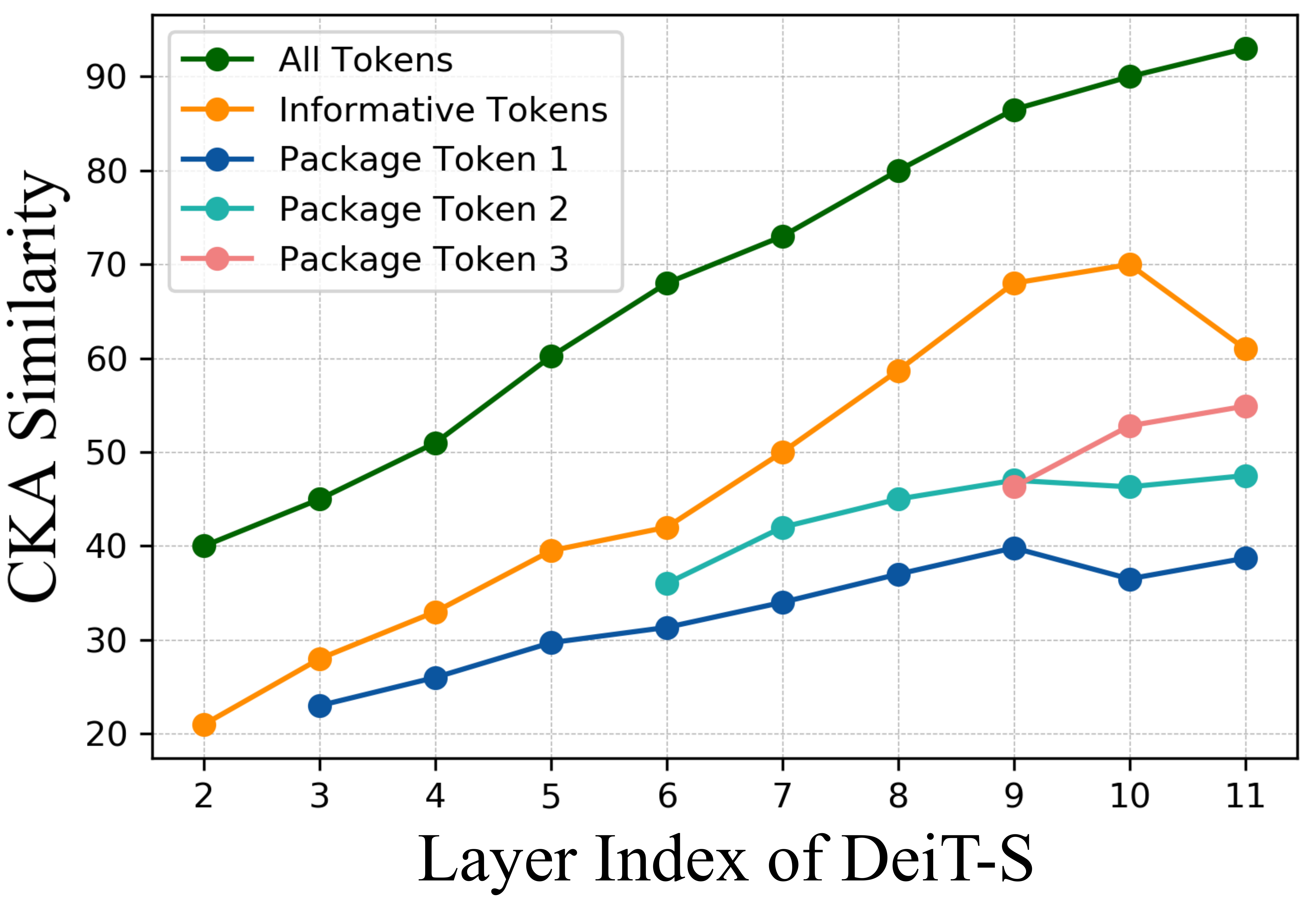}
\vspace{-0.1in}
\caption{CKA between the final CLS token and other tokens measured after each transformer block (2 to 11).} 
\label{fig:token}
\vspace{-0.2in}
\end{figure}

\noindent\textbf{Token redundancy viewed by different transformer blocks.}
Fig.~\ref{fig:token} gives the centered kernel alignment (CKA)~\cite{kornblith2019similarity} that represents the similarity between the tokens in each transformer block and the final CLS token, showing a tendency from weak to strong.
It tells us that it is inaccurate for each transformer block to encode or evaluate the image tokens, especially in the front blocks.
Thus, pruning rates for these front blocks should be lower to avoid ruling the informative tokens out.
This also inspires our token packager technique to provide chances to make up for pruning mistakes (Section~\ref{sec:token_packaging}).

\subsection{Overview of \M}

To develop a hardware-efficient image-adaptive token pruning framework, we first design an effective adaptive token pruning module (also known as \textbf{token selector}) according to the vision redundancy in ViTs (Section~\ref{sec:co-design}). This module includes an \textit{attention-based multi-head token classifier} and a \textit{token packager}, as shown in Fig.~\ref{fig:framework}. To improve the pruning rate and accuracy, the token classifier incorporates different token-level redundancies in multiple heads to more accurately classify tokens and the token packager consolidates (instead of discarding) non-informative tokens to one informative token to reserve information for later transformer blocks. To improve the hardware efficiency, we choose linear layers for the token selector to reuse the GEMM hardware component for the backbone ViT, and always concatenate the classified sparse tokens into dense ones.
Please note that it is challenging for CNN-based architecture to implement the data-level pruning,
because the kernel size of the convolution operation is fixed so that the irregular input features cannot be directly concatenated into dense ones to speedup the model inference.
Moreover, we deploy 8-bit fixed-point quantization to further compress the model, and propose a polynomial approximation of nonlinear functions inside ViTs for FPGA-based efficient implementations and regularize the quantization errors.
A proof-of-concept hardware accelerator design is presented in Section~\ref{sec:hardware_design}. 
To meet the target ViT inference speed on hardware while reserving the accuracy (typically within 1\% accuracy loss), one also needs to carefully decide the number of token selectors to insert in the backbone ViT, as well as the location and pruning rate of each token selector.
To address this challenge, in Section~\ref{sec:vit_training}, we propose a latency-aware multi-stage training strategy to learn all these parameters.



\section{Adaptive Token Pruning Module}
\label{sec:co-design}

\begin{figure*}[htb]
\centering
\includegraphics[width=1.9\columnwidth]{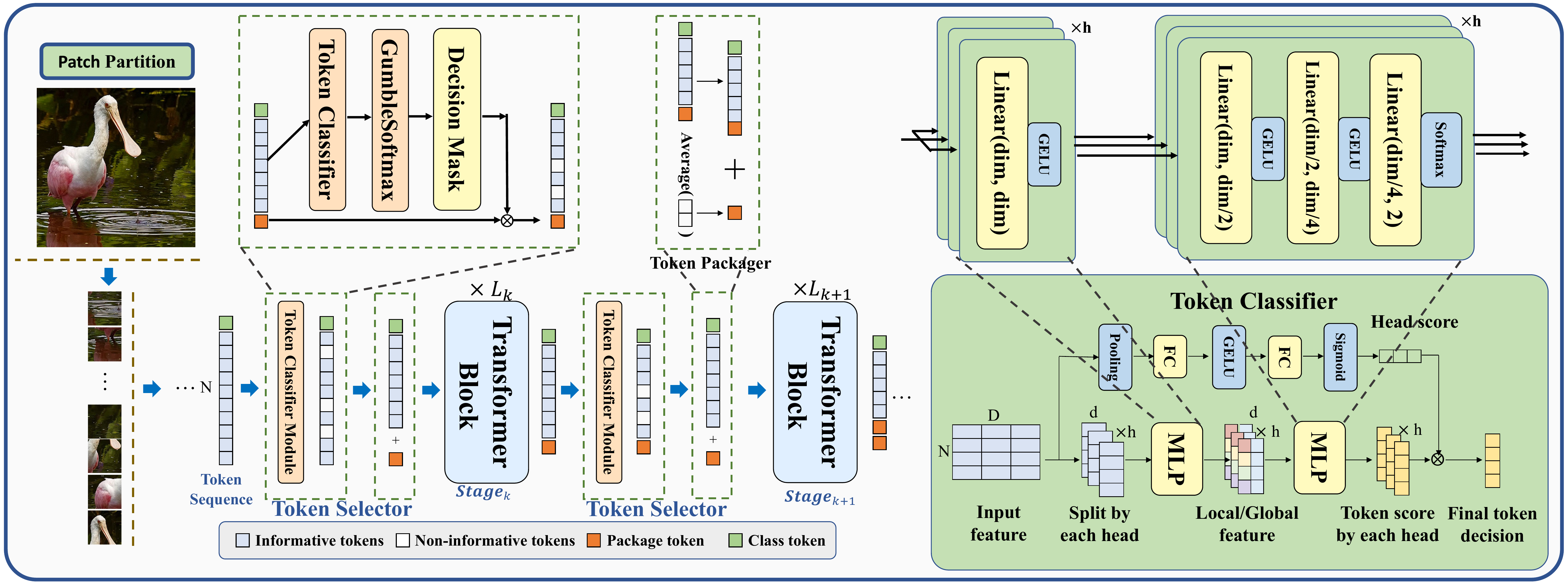}
\vspace{-8pt}
\caption{\M~overview. Left: Transformer blocks split into multiple stages with token selectors inserted between them (One token selector includes one token classifier and one token packager). Right: Multi-head token classifier to identify informative tokens.}
\label{fig:framework}
\vspace{-0.2in}
\end{figure*}

As shown in Fig.~\ref{fig:framework}, the proposed adaptive token selector is based on observations in Section~\ref{sec:motivation}, including the attention-based multi-head token classifier and the token packager.

\subsection{Attention-Based Multi-Head Token Classifier}
\label{sec:token_selector}

\noindent\textbf{Multi-Head Token Classifier.} 
Based on the multi-head visual receptive area of the ViT, we propose a fine-grained structure to evaluate token scores.
As shown in Fig.~\ref{fig:framework},  each head focuses on extracting different features and respective fields of an image, demonstrating that the importance of each token towards each head is different.
Hence, our multi-head classifier generates a score-map vector for each input token, marking the information amount  of each token in each head separately.

Let one head dimension be $d {=} D/h$.
We split the input $X {\in} \mathbb{R}^{N\times D}$ into $h$ subvectors $\{ x_i\}^{h}_{i=1}{\in} \mathbb{R}^{N\times d}$, and feed the subvectors into an MLP network to obtain the representation of local receptive field $E_i^\mathrm{local}$ and the representation of global receptive field $E_i^\mathrm{global}$.
\begin{equation}
E_i^\mathrm{local} = {\rm MLP}(x_i)  \in \mathbb{R}^{N\times d/2}
\label{eq:mlp1}
\end{equation}
\begin{equation}
E_i^\mathrm{global} = {\rm Average}( {\rm MLP}(x_i))  \in \mathbb{R}^{1\times d/2}
\label{eq:mlp2} 
\end{equation}
Pushing the feature $E_i {=} concate(E_i^{local},E_i^{global} \times N) {\in} \mathbb{R}^{N\times d}$ into another MLP network and one Softmax function, token score maps $\{s_i\}^{h}_{i=1}{\in} \mathbb{R}^{N\times 2}$ are produced. $s_i$ is the token score for attention head $i$ and ${\times} 2$ represents the keep and prune probabilities respectively:
\begin{equation}
\begin{gathered}
s_i = {\rm Softmax}({\rm MLP}(E_i)) \in \mathbb{R}^{N\times 2} \\
\end{gathered}
\end{equation}

\noindent\textbf{Attention-Based Branch.}
Base on the attention mechanism~\cite{hu2018squeeze}, we add an attention-based branch along the classifier backbone to synthesis the importance of each head:
\begin{equation}
\bar{X} = \mathrm{Concat}(\bigg\{ \frac{1}{d} \sum_{j=1}^{d}x_{ij} \bigg\}^{h}_{i=1}) \in \mathbb{R}^{N\times h}
\label{eq:avg_pool}
\end{equation}
\begin{equation}
A = {\rm Sigmoid} ({\rm MLP}(\bar{X})))  \in \mathbb{R}^{N\times h}
\vspace{-0.05in}
\label{eq:sigmoid_fc}
\end{equation}
where $\bar{X}$ is a head-wise statistic through its channel dimension $D$.
In Eq.~\eqref{eq:sigmoid_fc}, we feed $\bar{X}$ into an MLP network with the sigmoid to obtain the score vector $A$, evaluating the importance of each head.
Then the overall token score is calculated by the weighted average $S$ with $A$:
\begin{equation}
\tilde{S} = \frac{\sum_{i=1}^{h}s_i*a_i}{\sum_{i=1}^{h}a_i}   \in \mathbb{R}^{N\times 2}
\end{equation}
\vspace{-0.05in}
where $\tilde{S}$ is the token probability score. 
Then we apply the Gumbel-Softmax for the token keep/prune decision mask:

\vspace{-0.1in}
\begin{equation}
M={\rm GumbelSoftmax}(\tilde{S}) \in \{0,1 \}^N
\label{eq:gumbel}
\end{equation}
Since deleted image tokens cannot appear in subsequent blocks, $M$ passes on to the following blocks and will be updated by applying $M$ $\leftarrow$ $M\odot M^{\prime}$. $M^{\prime}$ is the new mask generated in the next stage.

\subsection{Token Packager}
\label{sec:token_packaging}

To solve the problem in the Sec~\ref{sec:motivation}, we apply a token packaging step that summarizes non-informative tokens (predicted by the classifier) into a package token instead of completely discarding them.
Assume there are $T$ (evaluated by the classifier) non-informative tokens $\{ \hat{x}_t\}^{T}_{t=1}{\in} \mathbb{R}^{T\times D}$ along with their token scores $\{ \tilde{s}_t\}^{T}_{t=1}{\in} \mathbb{R}^{T\times 2}$, these tokens are compressed into one token through weighted averaging:
\begin{equation}
P = \frac{ \sum_{t=1}^{T} \hat{x}_t * \tilde{s}_t[0]}{\sum_{t=1}^{T} \tilde{s}_t[0]} \in \mathbb{R}^{1\times D}
\end{equation}
where $P$ is the package token; $\tilde{s}_t[0]$ is the keep score of $\hat{x}_t$;
$*$ is an element-wise multiplication. 
$P$ will continue the subsequent calculations along with the informative ones, enabling the model to correct scoring mistakes.

After image-adaptive removing a certain part of tokens and the token packaging step, the sparse input matrix (all the informative tokens and package token) will be concatenated into a new smaller-size dense matrix to complete the computation in the following blocks, which speedup the model inference directly.
And most of the component operations (FC layer, Softmax, GELU) have already been there in the backbone ViT blocks. Therefore, we can utilize our unified operation-level optimization scheme of the ViT deployment on FPGA.

\section{ViT Hardware Accelerator Design with Adaptive Token Pruning} \label{sec:hardware_design}

The hardware architecture of ViT accelerators in \M~is illustrated in Fig.~\ref{fig:hardware}, including the accelerator design for pruned ViTs with token selector and the computation flow in the General Matrix Multiply (GEMM) engine. Besides the LayerNorm layer (less time consuming but more complex to implement on the FPGA) that is left on the ARM CPU, all other components in HeatVit are implemented on the FPGA. 


\subsection{Challenges}
To implement ViT FPGA accelerator with our dynamic token pruning, we address the following challenges.
(i) The token selector module should be implemented with minimal hardware overhead by adding miniature control logic and reusing existing hardware for the backbone ViT.
(ii) The GEMM loop tiling should accommodate an additional tiling dimension due to multi-head parallelism.
(iii) ViTs use more nonlinear operations than CNNs, and we need to refine these operations for more aggressive quantization and efficient hardware implementations without losing accuracy.

\subsection{ViT Accelerator with Dynamic Token Selection}
As displayed in Fig.~\ref{fig:AccHWwTokenSelector}, the input (tokens) and weight of each ViT layer are loaded from the off-chip DDR memory to the on-chip buffers, and processed by the GEMM engine which we will implement motivated by~\cite{fox2019training,lizh2022logic}.
Outputs are further processed with the activation function (Softmax or GELU), and then stored from the on-chip output buffer back to the off-chip memory.
Double buffering will be applied to overlap data transfer time with computation.
The token selector for pruning consists of FC layers and GELU activations, which also exist in the original ViTs, and thus can be managed by the same computation engine with negligible hardware overhead (Section~\ref{sec:token_selection}). Note LayerNorm is executed on the CPU.

\subsubsection{Loop Tiling in GEMM}
\label{sec:loop_GEMM}

\begin{figure}[tb]
\centering
\vspace{-0.2in}
\subfigure[ViT hardware accelerator architecture supporting dynamic token selection.
]{
\label{fig:AccHWwTokenSelector}
\begin{minipage}{0.5\textwidth}
\centering
\includegraphics[width=0.9\textwidth]{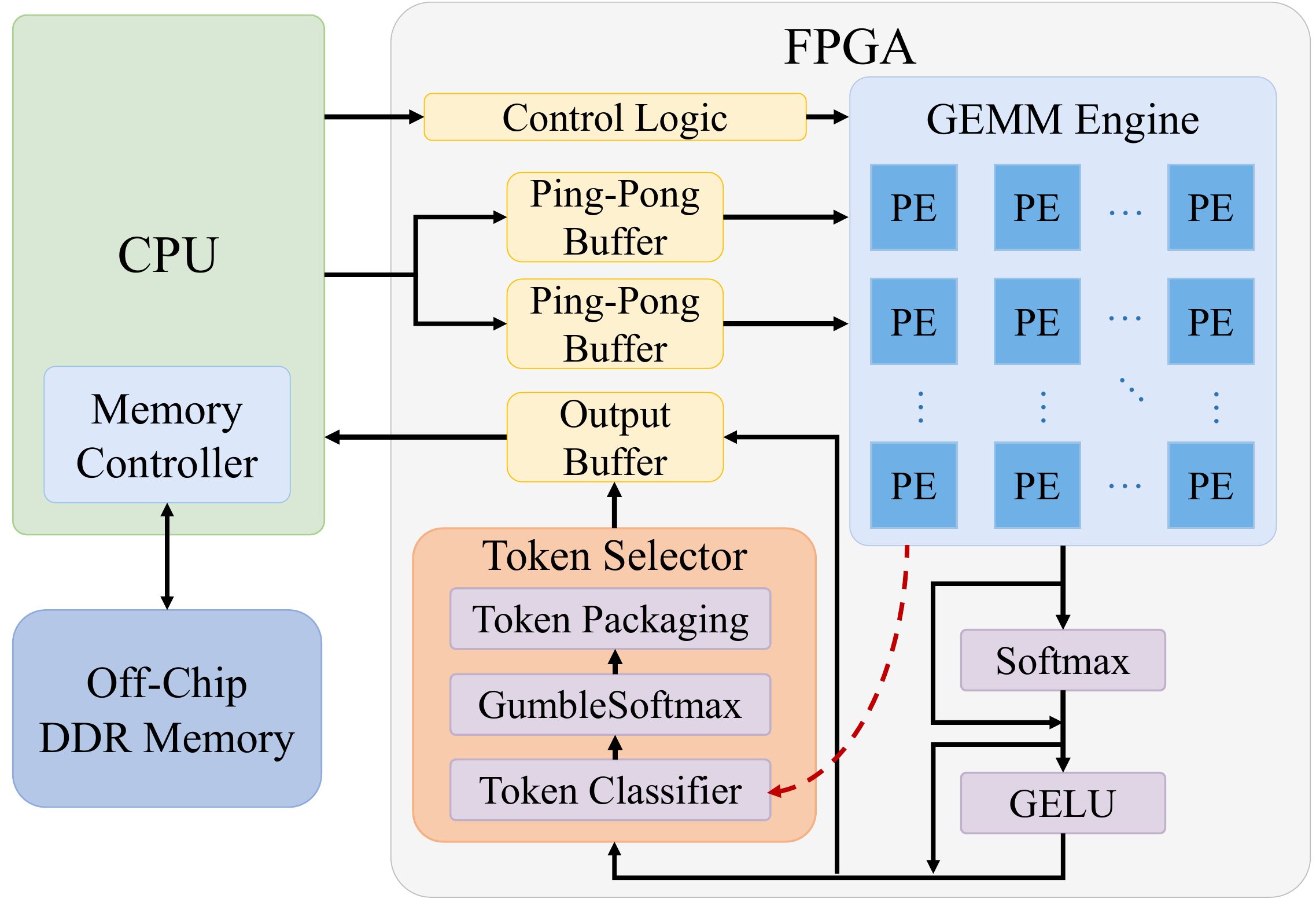}
\vspace{6pt} 
\end{minipage}
}
\hfill
\subfigure[GEMM loop tiling with an additional tiling from multi-head mechanism.]{
\label{fig:gemm_tiling}
\begin{minipage}{0.5\textwidth}
\centering
\includegraphics[width=0.9\textwidth]{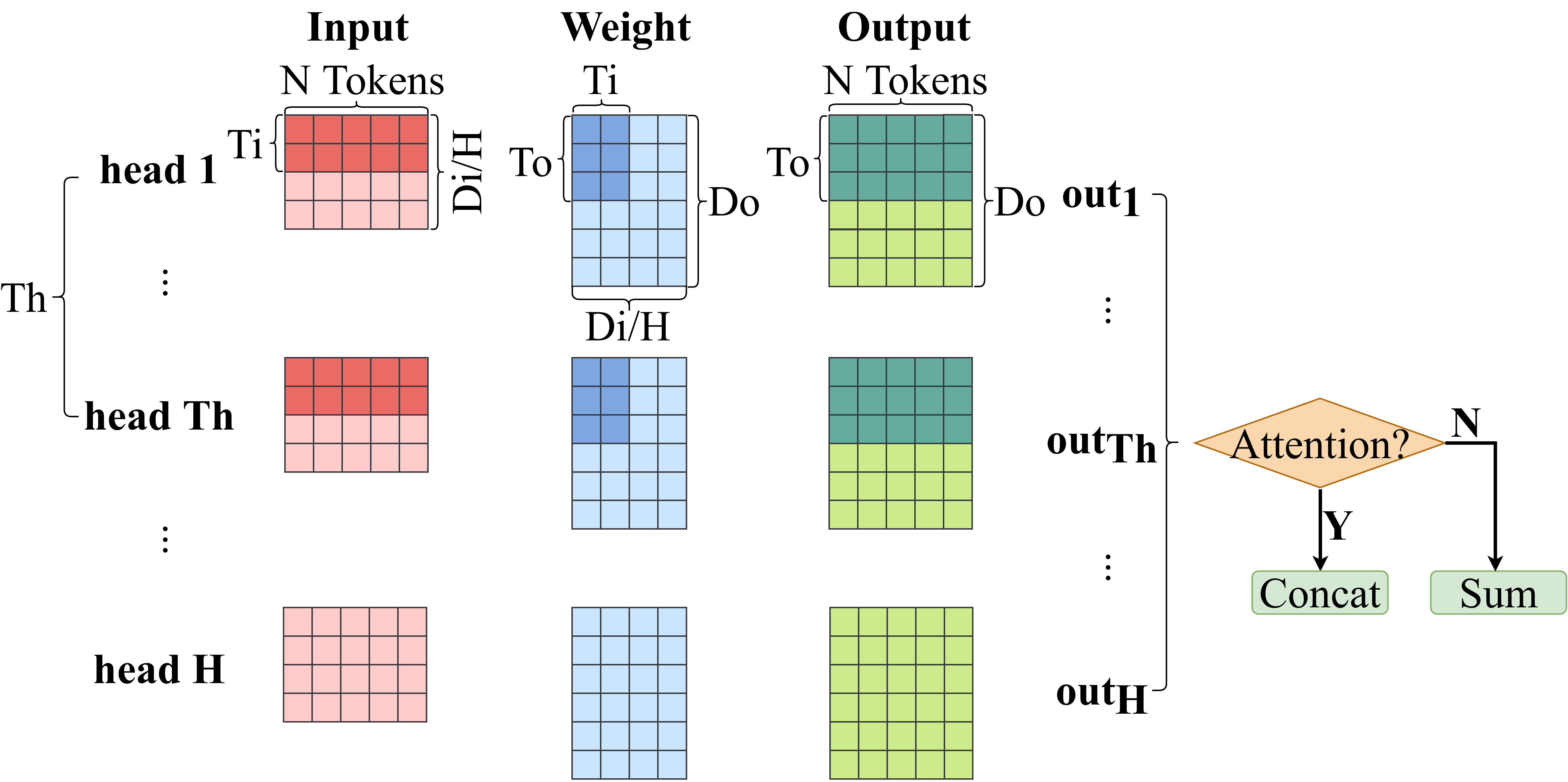}
\vspace{6pt}
\end{minipage}
}
\vspace{-10pt}
\caption{Hardware architecture of ViT accelerator in \M.}
\label{fig:hardware}
\vspace{-0.2in}
\end{figure}

We generalize loop tiling \cite{zhang2018caffeine} for GEMM engine to deal with relatively large ViT layers.
The concurrency of MAC operations in matrix multiplications requires pipelining and loop unrolling, as well as array partitioning of buffers. 
Fig.~\ref{fig:gemm_tiling} shows detailed computation flow in GEMM loop tiling for ViTs accommodating an additional tiling dimension from multi-head mechanism.
Given a ViT layer $j$ from \ding{172}$\sim$\ding{177} in Table~\ref{tab:complexity} that performs one or $h$ matrix multiplications with its $N_j$ tokens after pruning, we represent input size as $N_j \times D_i$, weight matrix size as $D_i \times D_o$, and  output size as $N_j \times D_o$, where $D_i$  denotes $D_{ch}$, $D_{attn}$, or $N$ in the MHSA module, and $D_{ch}$ or $4D_{fc}$ in the MLP module.
Tiling is applied to $D_i$ and $D_o$ dimensions, with tiling sizes  {$T_i$} and $T_o$, respectively. 
For attention computations ($\mathbf{Q} \times \mathbf{K}^\mathrm{T}$ in \ding{173} and $\mathbf{QK}^\mathrm{T} \times \mathbf{V}$ in \ding{174}), both input and output are split into $h$ groups. 
A control signal is used to indicate whether current computations are attention-related. 
Results from \ding{173} and \ding{174} are kept in $h$ groups, while results from other layers are accumulated.
To improve throughput, we optimize parallelism factors including $T_i$, $T_o$, and $T_h$ (an {additional tiling dimension for} $h$ heads). 
Therefore, we will conduct comprehensive FPGA resource modeling for available computing and on-chip memory resources.

\subsubsection{Throughput and Resource Utilization Analysis}

The inference throughput (FPS) of ViT inference is determined by the parallelism factors $T_i$, $T_o$ and $T_h$, which are bounded by the number of available computation (mainly DSPs) and on-chip memory (BRAMs) resources.
The inference is executed layer-by-layer, and the FPS can be inferred through dividing the total number of clock cycles required to process all the ViT layers with the working frequency on FPGAs.
The throughput and resource utilization analysis of ViT accelerators is similar to that of CNN accelerators~\cite{sun2022film-qnn}, except that the head dimension in ViTs needs to be additionally considered for parallelism.
Since the layers in token selectors can be managed by the same GEMM engine as for the existing ViT layers, this analysis also applies to token selectors.


\subsection{Token Selection Flow}
\label{sec:token_selection}
\begin{figure}[tb]
\centering
\includegraphics[width=0.65\columnwidth]{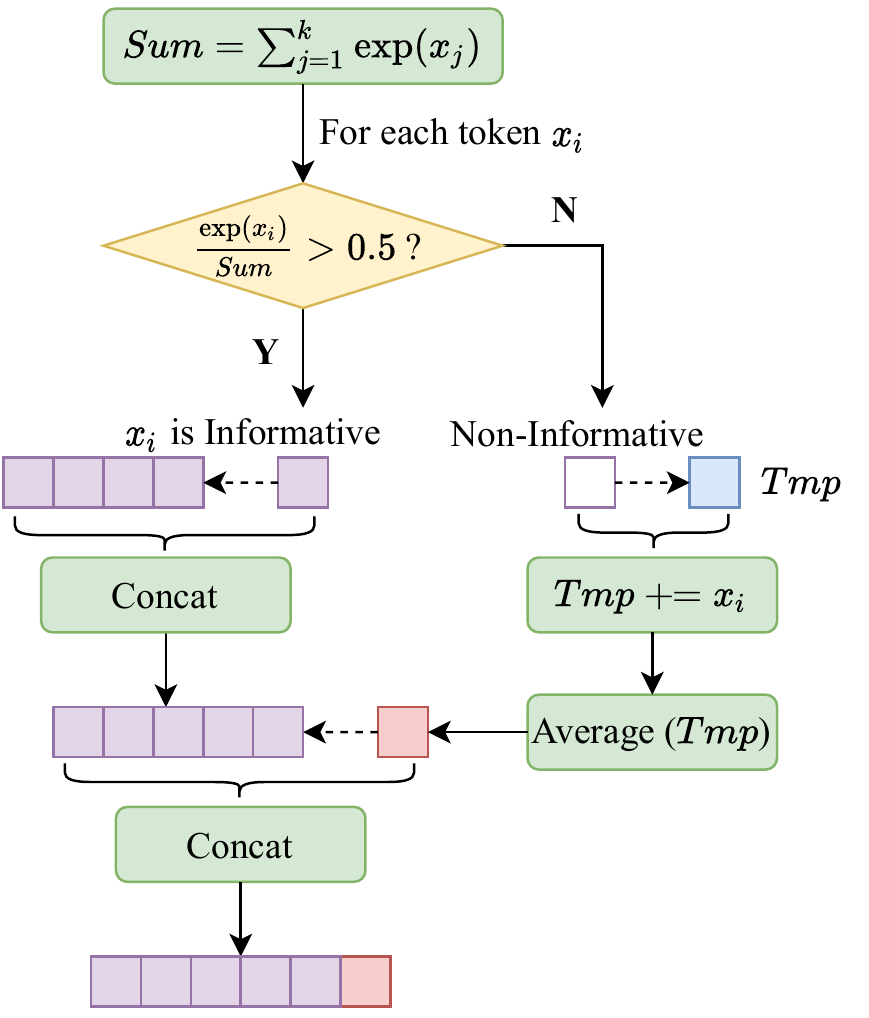}
\vspace{-0.1in}
\caption{Token selection flow.}
\label{fig:token_selector}
\vspace{-0.2in}
\end{figure}

Token selection contains token classification to determine informative and non-informative tokens using GumbelSoftmax with a threshold value (usually 0.5), and token packaging to average the non-informative tokens to one that is then consolidated into the informative tokens.
The pruned token (input) sequences with sparsity will be reorganized as dense ones, eliminating hardware overhead for indexing.
Fig.~\ref{fig:token_selector} describes the three steps to implement token selection in our ViT hardware accelerator: (1) calculating the exponent for each token $x_i$ and the summation $Sum$ of all these exponents;
(2) dividing each exponent by the sum and classifying the corresponding token as informative or not according to a threshold; and
(3) if the token is informative, concatenating it to the informative token sequence, otherwise adding it to a temporary token $Tmp$.
Finally, $Tmp$ is averaged and concatenated to the informative token sequence.

\subsection{Polynomial Approximation of Nonlinear Functions}
\label{sec:approx}
ViT models contain nonlinear functions including GELU, Softmax, and Sigmoid.
(i) Some nonlinear operations inside those functions, e.g., exponential function $\mathrm{exp}(x)$ and error function $\mathrm{erf}(x)$ consume large amounts of computing resources when implemented with the built-in Xilinx Vitis HLS math library \cite{vitis},  and thus incurring difficulty for hardware acceleration (Table~\ref{tab:nonlinearFunc_comp}).  
(ii) To apply more aggressive quantization than CNN/RNN models, we need to add the regularization effect on quantization error to these approximate operations.
Inspired by~\cite{gelu_softmax_approx}, we propose to explore algorithm-level polynomial approximation to implement GELU and Softmax functions, through which we introduce $\delta_1$ and $\delta_2$ (both $<$1) to control the regularization effect.
Since the Sigmoid function is only present inside token selectors (a small number), we do not introduce a regularization effect for it.

The $\mathrm{erf}(x)$ function is approximated using a second-order polynomial as,
\begin{equation}\label{eq:erf_polynomial}
    L_\mathrm{erf}(x) = \mathrm{sign}(x)\cdot\delta_1\cdot[a(\mathrm{clip}(|x|, \max=-b) + b)^2 + 1],
\end{equation}
with the constants $a=-0.2888$, $b=-1.769$ and $\delta_1<1$.

The GELU function is then expressed as
\begin{equation}\label{eq:gelu_approximation}
    \mathrm{GELU_{aprx}}(x) = \frac{x}{2} \left[ 1 + L_\mathrm{erf} \left(\frac{x}{\sqrt{2}} \right) \right],
\end{equation}

The Softmax function is approximated as
\begin{equation}\label{eq:softmax_approximation}
    \mathrm{Softmax_{aprx}}(\mathbf{x}_i) = \frac{\delta_2\exp (\tilde{x}_{i})}{\sum_{j=1}^{N}\exp (\tilde{x}_{j})},
\end{equation}
with $\tilde{x}_{i} = x_{i} - x_{\max}$, $x_{\max} = \max_{i}(x_i)$ and $\delta_2<1$. This subtraction ensures the numerical stability during the approximation calculation, and all inputs can be decomposed as $\tilde{x} = (-\ln2)z + p$, where $z$ is a non-negative integer and $p$ is a real number in $(-\ln2,0]$.
$\exp(\tilde{x})$ can then be calculated as $\exp(p)>>z$,
where $z = \left \lfloor -\tilde{x}/\ln2 \right \rfloor$, $p = \tilde{x} + z\ln2$, and $\exp(p)$ is approximated as
\begin{equation}\label{eq:exp_polynomial}
    \exp(p) = 0.3585(p + 1.353)^2 + 0.344.
\end{equation}
Both $\delta_1$ and $\delta_2$ are regularization value ($<1$) on quantization error, which can be constant ($\delta_1$=0.5, $\delta_2$=0.5 in our case).

For the Sigmoid function, we adopt the piece-wise linear approximation (PLAN) from~\cite{sigmoid_approx}.

\begin{table}[htb]
\centering
\tabcolsep 3pt
\vspace{-0.15in}
\caption{Resource utilization for nonlinear functions between original (Orig.) and approximation (Aprx.) implementations.}
\vspace{-0.1in}
\label{tab:nonlinearFunc_comp}

\begin{tabular}{c|cc|cc|cc}
\toprule
\multirow{2}{*}{} & \multicolumn{2}{c|}{GELU} & \multicolumn{2}{c|}{Sigmoid} & \multicolumn{2}{c}{Softmax} \\
\cline{2-7}
 &
  Aprx. &
  Orig. &
  Aprx. &
  Orig. &
  Aprx. &
  Orig. \\
\midrule
FF                & 334        & 191116       & 1015          & 2334         & 1939         & 2464         \\
LUT               & 438        & 160909       & 1512          & 2333         & 2364         & 2476         \\
DSP               & 4          & 139          & 0             & 3            & 2            & 3            \\ \bottomrule
\end{tabular}
\end{table}

Table~\ref{tab:nonlinearFunc_comp} compares the resource utilization for these nonlinear functions between the original implementations using the built-in Xilinx Vitis HLS math library and our proposed implementations.
Our methods are more resource efficient than those using the HLS math library, with 1.5$\times$$\sim$572$\times$ resource improvement. 
Furthermore, for each model, we try multiple sets of token pruning ratios and there is no accuracy drops between the approximate model and the original one.

\begin{figure}[tb]
\vspace{-0.1in}
\centering
\includegraphics[width=0.8\columnwidth]{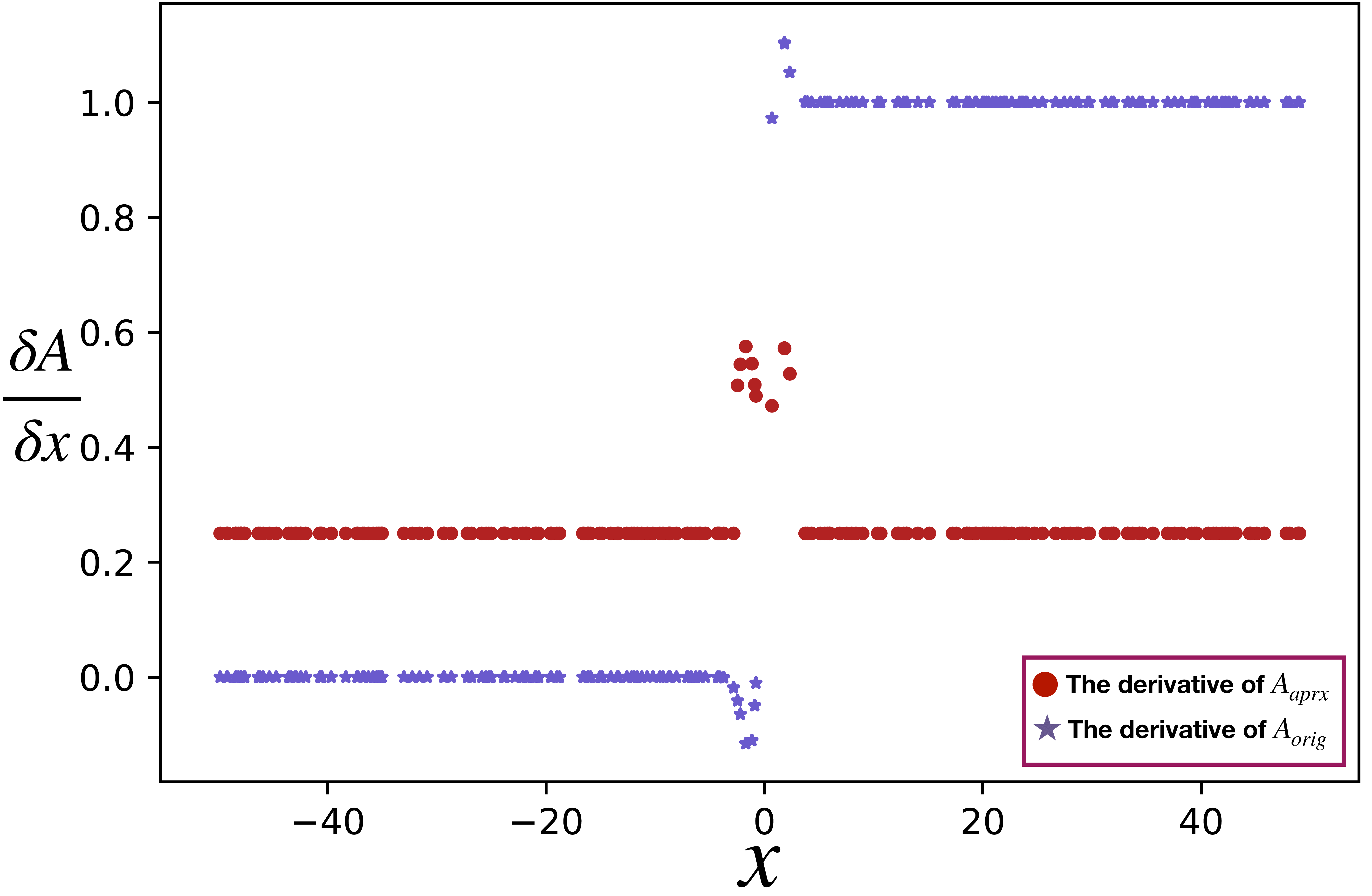}
\vspace{-10pt}
\caption{Regularization effect on quantization error of approximated GELU.}
\label{fig:approximated_gelu}
\vspace{-0.2in}
\end{figure}

\subsection{Regularization Effect on Quantization Error}
GELU and Softmax functions are abundant in transformer blocks, which inspires us to introduce the regularization effect of quantization error into the approximated functions for more aggressive quantization (e.g., 8-bit fixed-point quantization in our case). Here we proof that our regularization works.

GELU for activation data is: $A$$=$$\mathrm{GELU}(x)$.
Quantization on $x$ can be considered as adding a small error $\Delta e$ to $x$.
We examine the influence of $\Delta e$ on output $A$ by computing the GELU derivative $\frac{\partial A}{\partial x}$.
Assuming that $A$ changes by $\Delta e$$>$0, we can obtain the absolute error of output $A$:
\begin{equation}\label{eq:gelu_error}
    \mathrm{Error_{gelu}}(x) = \frac{\partial A}{\partial x} \cdot \Delta e.
\end{equation}
Since $\frac{\partial A_{aprx}}{\partial x}$ is always $<1$ (Fig.~\ref{fig:approximated_gelu}) for GELU, the total quantization error is reduced after approximation.

Softmax for activation data is: $A_i$$=$$\frac{\delta_2\exp (\tilde{x}_{i})}{\sum_{j=1}^{N}\exp (\tilde{x}_{j})}$.
As a similar process as GELU, we compute the $Softmax_{aprx}$ derivative $\frac{\partial A}{\partial x}$:
\begin{equation}\label{eq:softmax_derivative}
    \frac{\partial A}{\partial x} \! = \!
    \left\{\begin{matrix}
    \frac{\partial \frac{\delta_2\exp (\tilde{x}_{i})}{\sum_{j=1}^{N}\exp (\tilde{x}_{j})}}{\partial \tilde{x}_{j}}=\delta_2\cdot A_i\cdot (1-A_i), & i=j\\ 
    \frac{\partial \frac{\delta_2\exp (\tilde{x}_{i})}{\sum_{j=1}^{N}\exp (\tilde{x}_{j})}}{\partial \tilde{x}_{i}}=-\delta_2\cdot A_i\cdot A_j, & i\neq j   \end{matrix}\right. .
\end{equation}
Assuming $A_0$ changes by $\Delta e_0$, the absolute error of all outputs with Equation~(\ref{eq:softmax_derivative}) is:
\begin{equation}\label{eq:softmax_error}
\begin{aligned}
    Error_\mathrm{softmax} &= |\delta_2\Delta e_0 A_0 (1-A_0)|+\sum_{i=1}^{N-1}|-\delta_2\Delta e_0 A_0  A_i| \\
    &= 2\delta_2|\Delta e_0|\ A_0 (1-A_0) < \Delta e_0,
\end{aligned}
\end{equation}
since $0\leq A_0\leq 1$, $2A_0(1-A_0)$ is always smaller than 1 and $2\delta_2A_0(1-A_0)$ is further reduced ($\delta_2<1$). So, the total quantization error after Softmax approximation is $< \Delta e_0$.
\section{Latency-Aware Multi-Stage Training Strategy}\label{sec:vit_training}
Our strategy is as follows: (1) We propose latency-sparsity loss to take into account the latency characteristics of the hardware side during the training process.
(2) We design a block-to-stage training pipeline to learn the number of token selectors to insert in the backbone ViT, their location and pruning rates.
Please note that (i) the block-to-stage pipeline is based on token pruning, and 8-bit quantization on weight and activation will lead to 2$\times$$\sim$2.4$\times$ speedup without accuracy drops; and (ii) our training strategy uses finetuning for each selector, and the training effort of our pipeline is equivalent to the effort of training-from-scratch of the backbone ViT.


\noindent\textbf{Relationship between Latency and Keep Ratio.}
In order to bridge the inference of ViT model to the actual latency bound of hardware deployment, we build the latency-sparsity table for the target FPGA as Table~\ref{tab:speed_ratio}.
In this paper, we measure the actual numbers from our FPGA implementation.
Each time we only enter the $Keep Ratio$ tokens into one ViT block with a selector and test the corresponding latency.

\begin{table}[htb]
\centering
\tabcolsep 4pt
\vspace{-0.15in}
\caption{Tested latency of one DeiT block with different token keeping ratios on ZCU102 FPGA.}
\label{tab:speed_ratio}
\vspace{-0.1in}
\begin{tabular}{c|c|cccccc}
\toprule
\multicolumn{2}{c|}{\bf Keep Ratio}     & 1.0        & 0.9 & 0.8 & 0.7 & 0.6 & 0.5 \\
\midrule
\bf Latency & DeiT-T   & 1.034 & 0.945 & 0.881 & 0.764 & 0.702 & 0.636 \\
 (ms) & DeiT-S   & 3.161 & 2.837 & 2.565 & 2.255 & 1.973 & 1.682 \\
\bottomrule
\end{tabular}
\vspace{-0.1in}
\end{table}

\noindent\textbf{Latency-Sparsity Loss.} $\xi_{ratio}$ is built as follows:
\begin{equation}
{\mathrm{Block}(\rho_{i})}=latency\_sparsity\_table(\rho_{i})
    \label{eq:single_block}
\end{equation}
\begin{equation}
    \sum_{i=1}^{L}{\mathrm Block}_{i}(\rho_{i}) \leq LatencyLimit
    \label{eq:hardware_cost}
\end{equation}
\begin{equation}
    \xi_{ratio}=\sum_{i=1}^{L}(1-\rho _{i}-\frac{1}{B}\sum_{b=1}^{B}\sum_{j=1}^{N}D_{j}^{i,b})^{2}
    \label{eq:hd_loss}
\end{equation}
where Eq.~\eqref{eq:single_block} shows the look-up-table mapping  to find the latency of one $\mathrm{Block}$ under the corresponding ratio $\rho_i$ in Table~\ref{tab:speed_ratio}.
Eq.~\eqref{eq:hardware_cost} guarantees that the inference speed of the whole model satisfies the given hardware latency requirement,
with $i$ being the block index, 
$\rho_{i}$ being the corresponding pruning rate. 
Through Eq.~\eqref{eq:single_block} and \eqref{eq:hardware_cost}, we obtain appropriate $\rho_{i}$ and feed it to the latency-sparsity loss (\ref{eq:hd_loss}), where $B$ is the training batch size, and $M$ (Eq.~\eqref{eq:gumbel}) is the token keep decision.
In order to achieve per-image adaptive pruning, we set the average pruning rate of all images in one batch as the convergence target of the Eq.~\eqref{eq:hd_loss}.

\noindent\textbf{Training Objective.} It includes the standard cross-entropy loss $\xi_{cls}$, distillation loss $\xi_{distill}$, and latency-sparsity loss $\xi_{ratio}$.
The former two are the same as the loss strategy used in DeiT.
\begin{equation}
\xi = \xi_{cls}+\lambda _{distill}\xi_{distill}+\lambda _{ratio}\xi_{ratio}
    \label{eq:total_loss}
\end{equation}
where we set $\lambda_{distill}{=}0.5$, $\lambda_{ratio}{=}2$ in all our experiments.

\begin{algorithm}[htb]
\caption{Latency-Aware Multi-Stage Training with Image-Adaptive Token Pruning}
\label{alg:progressive_training}
\SetKwInOut{Input}{Input}
\SetKwInOut{Output}{Output}

\small
    \Input{ViT blocks $\{\mathrm{Block}_i\}^L_{i=1}$; \\
    \quad Accuracy drop constraint $a_\mathrm{drop}$; \\
    \quad Initial pruning rate $\rho_\mathrm{init}$; \\
    \quad Target latency $LatencyLimit$. \\}
    
    \Output{Token selectors with pruning rates $\rho_{s_1},...,\rho_{s_k}$. }
    \BlankLine
    
    \tcp{\scriptsize Step1: Insert a token selector between each two adjacent blocks and adjust the pruning rate $\rho_i$.}
    \ForEach{$i \in [L, L-1, \dots, 4]$}{
    $\rho_{i} = \rho_\mathrm{init}$\;
    $a,t \leftarrow$ Evaluate$(\{\mathrm{Block}_j(\rho_j)\}^L_{j=1})$\;
    \texttt{\scriptsize // $a$ and $t$ represent accuracy drop and latency of the whole model.} \\
    \While{$a < a_\mathrm{drop}$}{
    \eIf{$t<LatencyLimit$}{ Return the finalized token pruned ViT\;}
    {Decrease $t_i$\;
    $\rho_i$ = latency\_sparsity\_table($t_i$)\;
    $a,t \leftarrow$ Evaluate$(\{\mathrm{Block}_j(\rho_j)\}^L_{j=1})$;
    }}}
    \BlankLine
    
    \tcp{\scriptsize Step2: Combine sequential selectors with similar pruning rates as one stage, keep the first selector and retrain ViT. }
    
    $\rho_{s_1},...,\rho_{s_k} \leftarrow$ Combine $\rho_{1},...,\rho_{L}$;
    \BlankLine
    
    Retrain ViT$[\mathrm{Block}_1(\rho_1),..,\mathrm{Block}_{i}(\rho_{s_1}),..,\mathrm{Block}_L(\rho_{s_k})]$\;
    \eIf{$t<LatencyLimit$}
    {Return the finalized token pruned ViT\;}
    {Increase $a_{drop}$ or $LatencyLimit$\;
    Initialize the model and selectors from the end of the last Step1\;
    Go to the Step1 and repeat the training process.
    }
 
\end{algorithm}
\setlength{\textfloatsep}{8pt}
\vspace{-0.2in}

\begin{figure}[tb]
\centering
\vspace{-0.15in}
\includegraphics[width=0.9\columnwidth]{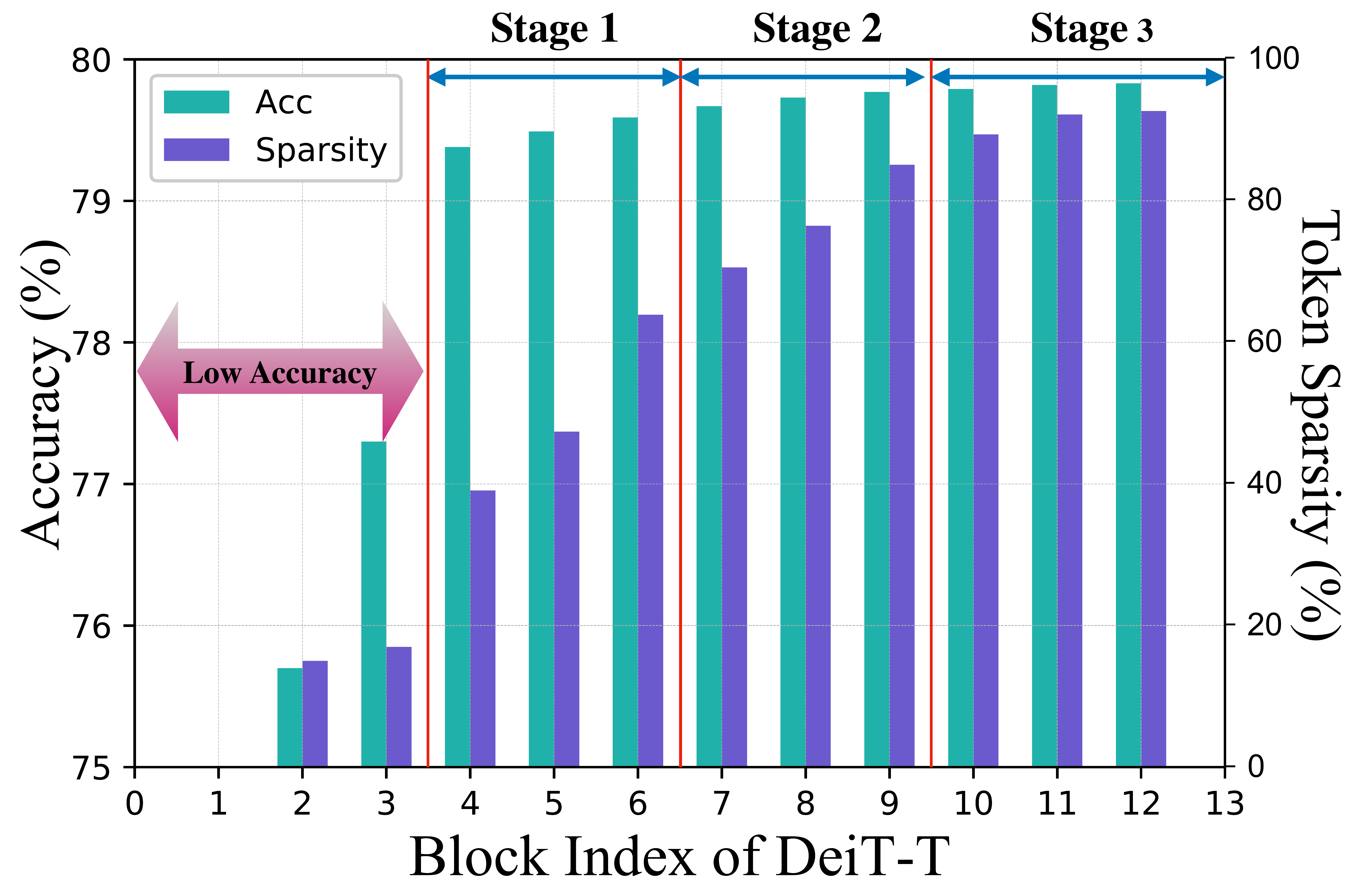}
\vspace{-0.15in}
\caption{Accuracy and token sparsity distribution after block-to-stage training. The insertion before the 2nd and 3rd blocks shows the pruning difficulty.}
\label{fig:model_location}
\vspace{-0.1cm}
\end{figure}

\noindent\textbf{Block-to-Stage Training.}
Algorithm~\ref{alg:progressive_training} presents our training strategy to find the optimal accuracy-pruning rate trade-offs and proper locations for token selectors, 
based on the token redundancy (Fig.~\ref{fig:token}).
In ViTs, tokens can be more effectively encoded in later blocks.
Hence, we adopt progressive training on inserting the token selector from later blocks to front ones.
Each time we insert a token selector, we train the current selector and fine tune the other parts by decreasing the latency of the current block until accuracy drops noticeably ($>0.5\%$).
Since token pruning in the front 3 blocks leads to more severe accuracy drops, we stop the selector insertion when reaching the fourth block.
If the pruned model has satisfied the target latency, we end the training and finalize the pruned model; otherwise, continue the training and repeat the insertion. After all the insertions, if the adjacent selectors have a similar pruning rate (difference$<8.5\%$), we combine them as one pruning stage and solely keep the first selector of that stage, shown in Fig.~\ref{fig:model_location}.
Finally, if the final latency of the whole model is lower than the target latency, we return the pruned model; otherwise, we will relax the accuracy or latency constraints and repeat the training.

\section{Experimental Results}
\label{sec:experiments}

\subsection{Experimental Setup}

Our experiments include adaptive token pruning and hardware implementation for ViTs with different pruning settings. Note that after token pruning we will apply 8-bit quantization on weight and activation, and all the quantization processes do not lose accuracy, except for 1.2\% on DeiT-T.

\subsubsection{Training Setup for ViT Pruning}

The baseline models with 32-bit floating-point precision are from the TorchVision library~\cite{Torchvision}.
Our experiments are conducted on the ImageNet-1K dataset with various transformers backbones, including DeiT-T, DeiT-S, DeiT-B, LV-ViT-S, and LV-ViT-M, as shown in Table~\ref{tab:vit_backbone}.
We follow the training settings in DeiT.
Training on one selector insertion costs 30 epochs on 8 NVIDIA A100-SXM4-40GB GPUs.
The training effort on block-to-stage training is listed in Table~\ref{tab:vit_backbone} which illustrates that the training effort of the entire block-to-stage pipeline is equivalent to the train-from-scratch of the backbone ViT.
Through our training pipeline, we observe that 3$\sim$4 token selectors are suitable for most of the models.

\begin{table}[htb]
\centering
\vspace{-0.1in}
\caption{Training effort for ViTs with different backbones.}
\label{tab:vit_backbone}
\vspace{-0.1in}
\begin{tabular}{c|c|c|c|cc}
\hline
\multirow{2}{*}{\textbf{Model}} & \multirow{2}{*}{\textbf{\#Heads}} & \multirow{2}{*}{\textbf{\begin{tabular}[c]{@{}c@{}}Embed. \\ Dim.\end{tabular}}} & \multirow{2}{*}{\textbf{Depth}} & \multicolumn{2}{c}{\textbf{\#Epochs for Training}}                   \\ \cline{5-6} 
                                &                                   &                                                                                  &                                 & \multicolumn{1}{c|}{\textbf{Baseline}} & \textbf{Ours} \\ \hline
DeiT-T                          & 3                                 & 192                                                                              & 12                              & \multicolumn{1}{c|}{300}               & 270                            \\ \hline
DeiT-S                          & 6                                 & 384                                                                              & 12                              & \multicolumn{1}{c|}{300}               & 270                            \\ \hline
DeiT-B                          & 12                                & 768                                                                              & 12                              & \multicolumn{1}{c|}{300}               & 270                            \\ \hline
LV-ViT-S                        & 6                                 & 384                                                                              & 16                              & \multicolumn{1}{c|}{400}               & 390                            \\ \hline
LV-ViT-M                        & 8                                 & 512                                                                              & 20                              & \multicolumn{1}{c|}{400}               & 390                            \\ \hline
\end{tabular}
\vspace{-0.2in}
\end{table}

\subsubsection{Hardware Platform}

Our hardware accelerator designs are evaluated on the Xilinx ZCU102 platform~\cite{ZCU102} with Zynq UltraScale+ MPSoC, containing 2520 DSPs, 912 BRAM blocks, and 274.1k LUTs.
We use Vitis HLS to generate the FPGA accelerators.
The synthesis and implementation of the designs are performed with the Xilinx Vitis 2020.1 tool~\cite{vitis}. The working frequency is set to 150 MHz.
The data in all models are represented in an 8-bit fixed-point format.
The hardware design for \M~is built based on state-of-the-art FPGA design for ViT~\cite{lizh2022logic} with the GEMM engine described in Section~\ref{sec:loop_GEMM}. Additionally, the \M~hardware design incorporates a token selector and polynomial approximation of nonlinear functions
explained in Section~\ref{sec:token_selection} and~\ref{sec:approx}.

\subsection{Accuracy and GMACs Results}

Fig.~\ref{fig:user-trade-off} demonstrates that our models achieve better accuracy-computation trade-offs compared to other pruned or scaled models.
Our \M~reduces the computation cost by $16.1\% {\sim} 42.6\%$ for various backbones with negligible $\leq 0.75\%$ accuracy degradation, which surpasses existing methods on both accuracy and efficiency.
To explore model scaling on ViT, we train more DeiT models with the embedding dimension of 160/256/288/320 as our baselines.
The accuracy improvement of HeatViT is 4\% (72.1\% vs. 68.1\% with 0.9 GMACs) over DeiT-T-160, 4.67\% (76.87\% vs. 72.20\% with 1.3 GMACs) over DeiT-T, and 0.81\% (79.34\% vs. 78.53\% with 2.64 GMACs) over DeiT-S-288.
Additionally, our method can prune up to 23.1\% on DeiT-T and 16.1\% on DeiT-S without any accuracy degradation.



\begin{figure}[tb]
\centering
\vspace{-0.2in}
\includegraphics[width=1 \columnwidth]{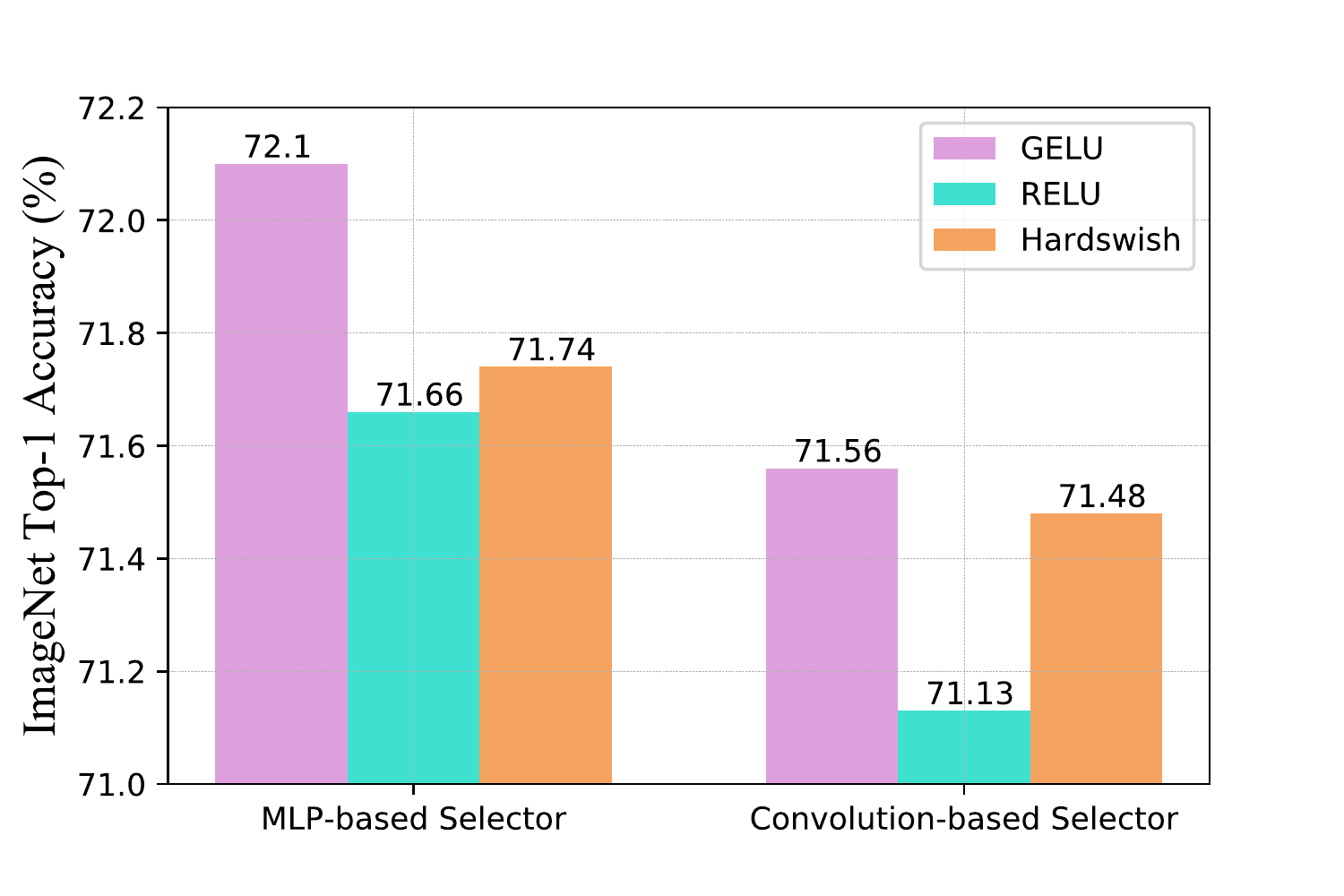}
\vspace{-0.3in}
\caption{Comparison of different token selector structures. We use DeiT-T (72.2\%, 1.3 GMACs) as the baseline.}
\label{fig:selector_structure}
\vspace{-0.1in}
\end{figure}

\begin{table*}[htb]
\centering
\tabcolsep 3pt

\vspace{0.05in}
\caption{Hardware results under different pruning settings for various ViTs on ImageNet dataset.}
\label{tab:result_hardware}
\vspace{-0.1in}
\begin{tabular}{c|c|ccc|cccc|c|cc}
\toprule
\multirow{2}{*}{\bf Design} & \multirow{2}{*}{\bf Model} & \bf Keep Ratio & \bf \#GMACs & \multirow{2}{*}{\bf Bitwidth} & \multicolumn{4}{c|}{\bf Resource Utilization} & \bf Power & \bf FPS & \bf Energy Effi. \\
\cline{6-9}
~ & ~ & (Stage 1/2/3) & (Pruning Rate) & ~ & kLUT & kFF & BRAM36 & DSP & (W) & (Accl. Rate) & (FPS/W) \\
\midrule
\multirow{4}{*}{Baseline} & DeiT-T & 1/1/1 & 1.30 (1$\times$) & 16 & \makecell{115.6 \\ (42\%)} & \makecell{101.5 \\ (19\%)} & \makecell{288.5 \\ (32\%)} & \makecell{1739 \\ (69\%)} & 8.012 & 78.3 (1$\times$) & 9.77 \\
\cline{2-12}

~ & DeiT-S & 1/1/1 & 4.60 (1$\times$) & 16 & \multirow{2}{*}{\makecell{130.3 \\ (48\%)}} & \multirow{2}{*}{\makecell{102.8 \\ (19\%)}} & \multirow{2}{*}{\makecell{492.5 \\ (54\%)}} & \multirow{2}{*}{\makecell{1754 \\ (70\%)}} & \multirow{2}{*}{10.095} & 25.9 (1$\times$) & 2.57 \\
\cline{2-5} \cline{11-12}

~ & LV-ViT-S & 1/1/1 & 6.55 (1$\times$) & 16 & ~ & ~ & ~ & ~ & ~ & 19.4 (1$\times$) & 1.92 \\
\cline{2-12}

~ & DeiT-B & 1/1/1 & 17.60 (1$\times$) & 16 & \makecell{144.5 \\ (53\%)} & \makecell{103.9 \\ (19\%)} & \makecell{664.3 \\ (73\%)} & \makecell{1786 \\ (71\%)} & 11.041 & 11.2 (1$\times$) & 1.01 \\

\midrule

\multirow{12}{*}{\makecell{\textbf{\M~} \\ with Token \\ Selector}} & \multirow{3}{*}{DeiT-T} & 0.85/0.79/0.51 & 1.00 (1.30$\times$) & 8 & \multirow{3}{*}{\makecell{137.6 \\ (50\%)}} & \multirow{3}{*}{\makecell{126 \\ (23\%)}} & \multirow{3}{*}{\makecell{355.5 \\ (39\%)}} & \multirow{3}{*}{\makecell{1968 \\ (78\%)}} & \multirow{3}{*}{\makecell{9.453}} & 183.4 (2.34$\times$) & 19.4  \\
\cline{3-5} \cline{11-12}
~ & ~ & 0.76/0.70/0.41 & 0.90 (1.44$\times$) & 8 & ~ & ~ & ~ & ~ & ~ & 198.8 (2.54$\times$) & 21.0 \\
\cline{3-5} \cline{11-12}
~ & ~ & 0.70/0.39/0.21 & 0.75 (1.74$\times$) & 8 & ~ & ~ & ~ & ~ & ~ & 271.2 (3.46$\times$) & 28.7 \\
\cline{3-5} \cline{11-12}
\cline{2-12}

~ & \multirow{3}{*}{DeiT-S} & 0.90/0.84/0.61 & 3.86 (1.19$\times$) & 8 & \multirow{6}{*}{\makecell{145 \\ (53\%)}} & \multirow{6}{*}{\makecell{100.4 \\ (18\%)}} & \multirow{6}{*}{\makecell{338.5 \\ (37\%)}} & \multirow{6}{*}{\makecell{1955 \\ (78\%)}} & \multirow{6}{*}{10.697} & 57.0 (2.20$\times$) & 5.33 \\
\cline{3-5} \cline{11-12}
\cline{3-5} \cline{11-12}
~ & ~ & 0.70/0.39/0.21 & 2.64 (1.74$\times$) & 8 & ~ & ~ & ~ & ~ & ~ & 97.1 (3.75$\times$) & 9.08 \\
\cline{3-5} \cline{11-12}
~ & ~ & 0.42/0.21/0.13 & 2.02 (2.27$\times$) & 8 & ~ & ~ & ~ & ~ & ~ & 109.2 (4.22$\times$) & 10.2 \\
\cline{2-5} \cline{11-12}

~ & \multirow{3}{*}{LV-ViT-S} & 0.90/0.84/0.61 & 5.49 (1.19$\times$) & 8 & ~ & ~ & ~ & ~ & ~ & 62.8 (3.24$\times$) & 5.87 \\
\cline{3-5} \cline{11-12}
~ & ~ & 0.70/0.39/0.21 & 3.77 (1.74$\times$) & 8 & ~ & ~ & ~ & ~ & ~ & 72.8 (3.75$\times$) & 6.81 \\
\cline{3-5} \cline{11-12}
~ & ~ & 0.42/0.21/0.13 & 2.88 (2.27$\times$) & 8 & ~ & ~ & ~ & ~ & ~ & 89.1 (4.59$\times$) & 8.33 \\

\cline{2-12}
~ & \multirow{3}{*}{DeiT-B} & 0.90/0.84/0.61 & 14.79 (1.19$\times$) & 8 & \multirow{3}{*}{\makecell{161.4 \\ (59\%)}} & \multirow{3}{*}{\makecell{101.8 \\ (19\%)}} & \multirow{3}{*}{\makecell{528.6 \\ (58\%)}} & \multirow{3}{*}{\makecell{2066 \\ (82\%)}} & \multirow{3}{*}{11.352} & 36.1 (3.22$\times$) & 3.18 \\
\cline{3-5} \cline{11-12}
~ & ~ & 0.70/0.39/0.21 & 10.11 (1.74$\times$) & 8 & ~ & ~ & ~ & ~ & ~ & 43.3 (3.87$\times$) & 3.81 \\
\cline{3-5} \cline{11-12}
~ & ~ & 0.42/0.21/0.13 & 7.75 (2.27$\times$) & 8 & ~ & ~ & ~ & ~ & ~ & 54.8 (4.89$\times$) & 4.83\\

\bottomrule
\end{tabular}
\vspace{-0.2in}
\end{table*}

\subsubsection{Operations in Token Selector}
We compare different selector structures with the same computation cost of 0.9 GMACs in Fig.~\ref{fig:selector_structure}.
MLP-based token selectors outperform convolution-based ones under different combinations. 
In addition, MLP-based selectors bring more benefits than convolution-based ones, since we reuse the GEMM on the FC operation of original ViTs without opening additional resources for the new computation design.

For the activation functions, GELU outperforms ReLU~\cite{agarap2018deep} and Hardswish~\cite{howard2019searching} continuously.
Even though ReLU and Hardswish can be directly deployed on FPGAs, the resource utilization of GELU can be improved to $35\times$$\sim$$572\times$ by the polynomial approximation (Section~\ref{sec:approx}).

\subsection{Hardware Results} 

Multiple hardware accelerators are designed according to the number of heads in a specific ViT. As shown in Table~\ref{tab:result_hardware}, with the same total degree of computation parallelism, the resource utilization and power of DeiT-S and LV-ViT-S designs are higher than those of DeiT-T ones, since DeiT-T has 3 heads while DeiT-S and LV-ViT-S all have 6 heads, requiring more BRAM space to accommodate data of all the attention heads. This trend is similar for DeiT-B (12 heads).

Compared with the baseline hardware designs (16-bit and no token pruning), the accelerators with token selector in \M~framework utilize 9\% more DSPs and 8\% more LUTs for DeiT-T, 8\% more DSPs and 5\% more LUTs for DeiT-S and LV-ViT-S, 11\% more DSPs and 6\% more LUTs for DeiT-B.
This demonstrates that the control flow to support adaptive token pruning introduces negligible resource utilization overhead.
After the token pruning, the frame rate increases from 78.3 FPS to 142.7 FPS (1.82$\times$) for DeiT-T,  from 25.9 FPS to 57.6 FPS (2.22$\times$) for DeiT-S, from 19.4 FPS to 46.9 FPS (2.42$\times$) for LV-ViT-S, and from 11.2 FPS to 28.9 FPS (2.58$\times$) for DeiT-B.
Furthermore, we deploy the 8-bit fixed-point quantization on models to achieve another 1.90$\times$ speedup, ending up the final speedup with 3.46$\times$ (271.2 FPS) for DeiT-T, 4.22$\times$ (109.2 FPS) for DeiT-S, 4.59$\times$ (89.1 FPS) for LV-ViT-S, and 4.89$\times$ (54.8 FPS) for the DeiT-B.

\subsubsection{Comparisons with CPUs and GPUs}



\begin{figure}[tb]
\centering
\includegraphics[width=1\columnwidth]{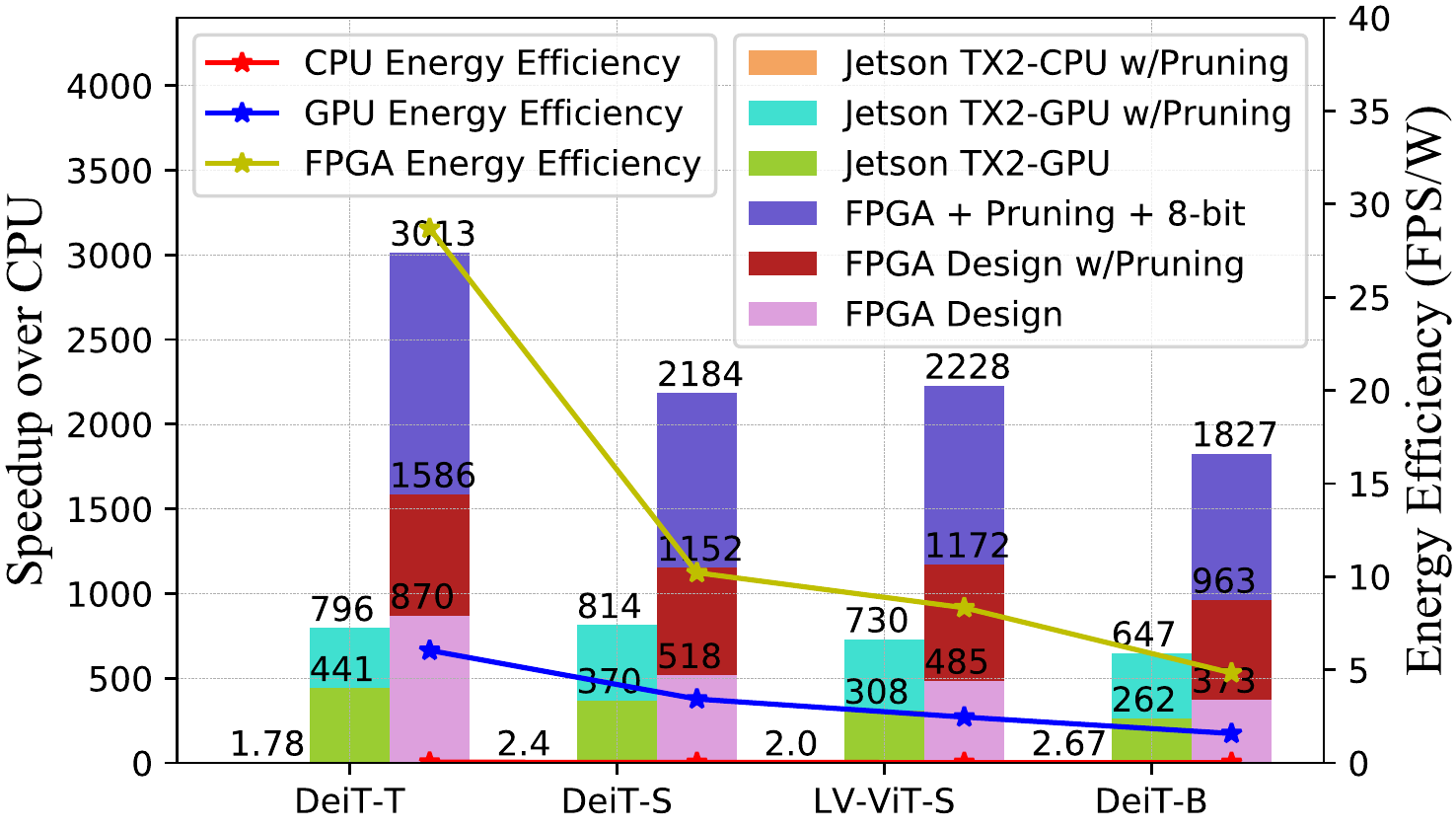}
\vspace{-0.3in}
\caption{Comparison of energy efficiency between HeatViT and TX2 CPU/GPU with the improvement breakdown of different techniques.}
\label{fig:comparison}
\vspace{-0.1in}
\end{figure}

We also test DeiT-T, DeiT-S, LV-ViT-S, and DeiT-B on Jetson TX2 with 4-core ARM CPU and NVIDIA Pascal GPU, and compared them with our FPGA (ZCU102) implementation.
Since MSA and FFN computations are reduced by token pruning, CPUs and GPUs can also be accelerated.
And TX2 CPU/GPU does not support low-bit computation,  so we only present the full precision model for them with adaptive token pruning as shown in Fig.~\ref{fig:comparison}.
All the results are normalized against the original model on TX2 CPU without token pruning. 
First, our final FPGA implementation achieves the highest $1827\times\sim3013\times$ speedup with 9.453W, 10.697W, and 11.352W power for different designs. 
While the baseline FPGA design~\cite{lizh2022logic} (16-bit and no token pruning) achieves 373$\times$$\sim$870$\times$ speedup, token pruning can bring 1.82$\times$$\sim$2.58$\times$ speedup and ambitious 8-bit quantization can contribute another 1.90$\times$ speedup.
Second, with token pruning, TX2 GPU achieves 647$\times$$\sim$814$\times$ speedup with 12W power and TX2 CPU achieves 1.78$\times$$\sim$2.67$\times$ speedup with 4W power.
For the energy efficiency, our FPGA implementation achieves 4.8 FPS/W$\sim$28.7 FPS/W, which is 242.6$\times$$\sim$719.0$\times$ higher than TX2 CPU and 3.0$\times$$\sim$4.7$\times$ higher than TX2 GPU (with token pruning).

\section{Conclusion}
In this paper, we have proposed a hardware-efficient image-adaptive token pruning framework called \M~for ViT inference acceleration on resource-constraint edge devices. 
To improve the pruning rate and accuracy, we analyzed the inherent computational patterns in ViTs and designed an effective token selector that can more accurately classify tokens and consolidates non-informative tokens.
We also implemented a proof-of-concept ViT hardware accelerator on FPGAs by heavily reusing the hardware components built for the backbone ViT to support the adaptive token pruning module.
Besides, we propose a polynomial approximation of nonlinear functions for ambitious (8-bit) quantization and efficient hardware implementation.
Finally, to meet both the target inference latency and model accuracy, we proposed a latency-aware multi-stage training strategy to learn the number of token selectors to insert into the backbone ViT, and the location and pruning rate of each token selector. Experimental results demonstrate that \M~achieves superior pruning rate and accuracy compared to state-of-the-art pruning studies, while incurring trivial amount of hardware resource overhead.

\section*{Acknowledgements}
This work is partly supported by NSF CCF-1901378;
NSERC Discovery Grant RGPIN-2019-04613, DGECR-2019-00120, Alliance Grant ALLRP-552042-2020;
CFI John R. Evans Leaders Fund.



\bibliographystyle{IEEEtranS}
\bibliography{refs}

\end{document}